\documentclass[journal]{IEEEtran}
\usepackage{fancyhdr}
\usepackage{algorithm}
\usepackage{algpseudocode}
\usepackage{fancybox}
\usepackage{times,color}
\usepackage{amsmath}
\usepackage{amsthm}
\usepackage{amssymb}
\usepackage[dvips]{graphicx}
\usepackage{bm}
\usepackage{subfigure}
\usepackage{threeparttable}
\usepackage{setspace}
\usepackage{booktabs}
\usepackage{subfigure}
\usepackage{mathrsfs}
\usepackage{stfloats}
\usepackage{diagbox}
\usepackage[numbers,sort&compress]{natbib}
\usepackage{array}
\usepackage{flushend}
\usepackage{color}

\ifCLASSINFOpdf
\else
\fi
\begin{document}
%
\title{Ziv-Zakai Bound for DOAs Estimation}
%
%
%

\author{Zongyu~Zhang,~\IEEEmembership{Student Member,~IEEE,}
        Zhiguo~Shi,~\IEEEmembership{Senior Member,~IEEE,}
        and~Yujie~Gu,~\IEEEmembership{Senior Member,~IEEE}
\thanks{Z. Zhang is supported by the China Scholarship Council for his stay at the University of Pisa. The work of Z. Zhang and Z. Shi is supported in part by National Natural Science Foundation of China (No. U21A20456, 61901413, 61772467). (Corresponding authors: Zhiguo Shi and Yujie Gu).}
\thanks{Z. Zhang is with the College of Information Science and Electronic Engineering, Zhejiang University, Hangzhou 310027, China, and also with the Department of Information Engineering, University of Pisa, Pisa, PI 56122, Italy (email: zongyu\_zhang@zju.edu.cn).}
\thanks{Z. Shi is with the College of Information Science and Electronic Engineering, Zhejiang University, Hangzhou 310027, China (email: shizg@zju.edu.cn).}
\thanks{Y. Gu is with Advanced Safety \& User Experience, Aptiv, Agoura Hills, CA 91301, USA (email: guyujie@hotmail.com).}}

\maketitle

\begin{abstract}
Lower bounds on the mean square error (MSE) play an important role in evaluating the direction-of-arrival (DOA) estimation performance.
Among numerous bounds for DOA estimation, the local Cram$\mathrm{\acute{e}}$r-Rao bound (CRB) is only tight asymptotically.
By contrast, the existing global tight Ziv-Zakai bound (ZZB) is appropriate for evaluating the single source estimation only.
In this paper, we derive an explicit ZZB applicable for evaluating hybrid coherent/incoherent multiple sources DOA estimation.
It is first shown that, a straightforward generalization of ZZB from single source estimation to multiple sources estimation cannot keep the bound valid in the \textit{a priori} performance region. To derive a global tight ZZB,
we then introduce order statistics to describe the change of the \textit{a priori} distribution of DOAs caused by ordering process during the MSE calculation.
The derived ZZB is for the first time formulated as a function of coherent coefficients between coherent sources, and reveals the relationship between the MSE convergency in the \textit{a priori} performance region and the number of sources. Moreover, the derived ZZB also provides a unified tight bound for both overdetermined and underdetermined DOA estimation.
Simulation results demonstrate the obvious advantages of the derived ZZB over the CRB on evaluating and predicting the estimation performance for multiple sources DOA.
\end{abstract}
%
\IEEEpeerreviewmaketitle
\section{Introduction}
\IEEEPARstart{D}{irection}-of-arrival (DOA) estimation is a fundamental problem in many array processing applications including radar, sonar, navigation, and wireless communications \cite{van2002detection}. In the past decades, multiple sources DOA estimation has attracted great research interest, the majority of which concentrated on the algorithm design. Among them, the typical ones include multiple signal classification (MUSIC) \cite{schmidt1986multiple, rao1989performance,liu2015remarks}, estimation of signal parameters via rotational invariant techniques (ESPRIT) \cite{roy1989esprit,Han2005An,Chen2010ESPRIT}, sparse reconstruction-based algorithms \cite{malioutov2005sparse,Zhou2018Direction,Zhou2018Offgrid}, and machine learning-based algorithms \cite{Elbir2020DeepMUSIC,Barthelme2021AMachine,Papageorgiou2021Deep}.

Mean square error (MSE) is commonly used to evaluate the performance of estimators. However, as a typical nonlinear parameter estimation problem, there is no exact closed-form minimum MSE for DOAs estimation, which motivates to find MSE lower bounds (see, for example, \cite{van2007Bayesian} and the references therein). Among them, Cram$\mathrm{\acute{e}}$r-Rao bound (CRB) \cite{Cramer2016Mathematical} is the most widely used one, since it is easily derived from the inverse of the Fisher information matrix. Therein, the closed-form CRB for DOAs estimation is given in \cite{Stoica1989MUSIC, Stoica1990Performance}.
To overcome the singularity of Fisher information matrix in the underdetermined estimation, nonsingular Fisher information matrix is derived in \cite{Wang2016Coarrays, Liu2017Cramer}, where the corresponding coarray CRB for underdetermined DOA estimation is also provided.
As a local non-Bayesian bound, it is well known that these CRBs are only asymptotically tight in small error estimation scenario (e.g., high signal-to-noise ratio (SNR)), and cannot offer a valid bound in evaluating estimation performance under low SNR situation.
In order to overcome the locality of these CRBs, numerous Bayesian bounds utilizing the \textit{a priori} distribution of parameters had been proposed in past few decades \cite{van2007Bayesian}, among which one typical class is covariance inequality-based bounds \cite{Renaux2008Afresh} including Bayesian CRB (BCRB) \cite{van2002detection}, Weiss-Weinstein bound (WWB) \cite{Weiss1985Alower}, Reuven-Messer bound (RMB) \cite{Reuven1997ABarankin}, Bayesian Abel bound (RAB) \cite{Renaux2006TheBayesian} and Bobrovsky-Zakai bound (BZB) \cite{Bobrovsky1976Alower}. However, these bounds rely on higher order derivative, optimization over free variables and empirical testing point selection to provide a global tighter solution \cite{van2007Bayesian}.

As another typical Bayesian bound, Ziv-Zakai bound (ZZB) enables to provide a global tight bound on the MSE over a wide range of SNR.
ZZB was originally proposed in \cite{Ziv1969Some} (with improvement in \cite{Seidman1970Performance, Chazan1975Improved}) by considering a scalar estimation problem, where the time delay is assumed to follow a uniform \textit{a priori} distribution. Since then, the ZZBs for specific scalar parameter estimation problems have been widely studied \cite{Gu1991Amodified,Sadler2010Ziv, KHAN2013Explicit, Decarli2014Ziv, Chiriac2015Ziv, Mishra2017Performance,Laas2021Onthe,Zhang2022Ziv}, e.g., single source DOA estimation exploiting linear array \cite{Gu1991Amodified, KHAN2013Explicit}.
On the other hand, a unified expression of ZZB for vector parameter estimation with arbitrary \textit{a priori} distribution was first derived in  \cite{Bell1997Extended}, which provides a promising framework for the derivation of ZZB on multiple sources DOA estimation.
However, it remains challenges in deriving an explicit ZZB due to the involved high dimensional integration and constrained optimization.
Although an explicit ZZB for two-dimensional DOA estimation was derived in \cite{Bell1996Explicit}, the involved Sherman-Morrison formula and matrix determinant lemma make it appropriate for single source estimation only.
Furthermore, its uniform distribution assumption on $\sin\theta$ rather than on the DOA $\theta$ itself, is impractical for DOA estimation, because it implies that, the source signal impinges on the array from the boresight with the highest probability, and the probability gradually decreases with the DOA  deviating from the boresight.
Hence, the existing ZZBs for single source DOA estimation are difficult to be straightforwardly extended to multiple sources scenario, where the main challenges are:
\vspace{-2pt}
\begin{itemize}
\item With the number of sources increasing, the numerical solution suffers from heavier computational burden, such that an explicit ZZB is required.
\item For multiple sources DOA estimation, there naturally exists permutation ambiguity during MSE calculation. However, the ordering process for the elimination of permutation ambiguity implicitly changes the \textit{a priori} distribution of DOAs, which makes the ZZB derived from the existing framework invalid in the \textit{a priori} performance region.
\item For multiple sources DOA estimation, the mutual coherence among coherent sources is generally unavoidable due to multipath propagation (see, \cite{Yang2019Source, Zheng2020Direction, Zheng2022Structured} and the references therein), the effect of which on the ZZB is still unclear.
\end{itemize}

\vspace{-2pt}
Due to these difficulties, some researchers had to adopt the ZZB for single source DOA estimation \cite{Bell1996Explicit} to evaluate their multiple sources DOA estimation \cite{Gupta2019Design}. In this paper, we derive an explicit ZZB for hybrid coherent/incoherent multiple sources DOA estimation.
By utilizing Woodbury matrix identity and Sylvester's determinant theorem, we first follow the derivation in \cite{Bell1996Explicit} to generalize the ZZB for single source DOA estimation to multiple sources DOA estimation.
However, such straightforward generalization does not consider the permutation ambiguity arising in MSE calculation for multiple sources DOA estimation, which makes the generalized ZZB invalid in the \textit{a priori} performance region. To this end, we for the first time introduce the order statistics to describe the effect of the elimination of permutation ambiguity on the ZZB, such that the ZZB keeps tight for the evaluation of DOA estimators outside the asymptotic region.
The ZZB for DOAs estimation is an explicit function of the number of sources, the number of array sensors, the number of snapshots, the \textit{a priori} distribution and SNRs of sources, array observation data, and the coherent coefficient.
Furthermore, the derived ZZB provides a unified expression for both overdetermined DOA estimation and underdetermined DOA estimation, and also explicitly reveals the MSE convergency in the \textit{a priori} performance region with respect to (w.r.t.) the number of sources and their \textit{a priori} distributions.
Simulations demonstrate the validity of the ZZB to predict multiple sources DOA estimation performance, benefited from its obvious threshold effect.
Compared with the widely used CRB, the derived ZZB provides a tight bound over a wider range of SNR, and clearly reveals the MSE convergency in the \textit{a priori} performance region and the effect of the coherent coefficient in the transition region.

The notations used in this paper are summarized in Table \ref{Table: Notations}.
The rest of this paper is organized as follows. We introduce a hybrid coherent/incoherent multiple sources model, MSE and CRB in Section \ref{Signal Model}, and the related ZZB preliminaries in Section \ref{ZZB Preliminaries}. Then, in Section \ref{ZZB derivation}, we derive a ZZB as an explicit function of multiple sources, where the permutation ambiguity in DOAs estimation is also considered. We perform simulations in Section \ref{Simulations} to demonstrate the advantages of the ZZB. Finally, we make our conclusions in Section \ref{Conclusions}.

\begin{table}[!ht]
\small
\centering
	\caption{List of Notations}
    \vspace{-5pt}
	\begin{spacing}{1.2}
    \begin{tabular}{c|l}
        \toprule
        Notation  & Description \\ \hline	
        $\mathbb{R}$ & Set of real numbers\\
		$\mathbb{C}$ & Set of complex numbers \\
		$[\, \cdot \, ]^{\mathrm{T}}$ & Transpose\\
		$[\, \cdot \, ]^{\mathrm{H}}$ & Hermitian transpose \\
		$\mathrm{Tr}\{ \, \cdot \, \}$ & Trace of a matrix\\
        $| \, \cdot \, |$ & Determinant of a matrix \\
        $\bm{0}_{M}$ & $M$-dimensional all-zero vector\\
        $\bm{1}_{M}$ & $M$-dimensional all-one vector\\
        $\bm{I}_{M}$ & $M$-dimensional identity matrix \\
        $\mathbb{E}\{\, \cdot \, \}$ & Statistical expectation \\
        $\mathrm{diag}[\,\cdot\,]$ & Diagonal matrix\\
        $\mathrm{block \ diag}[\,\cdot\,]$ & Block diagonal matrix\\
        $\|\, \cdot \,\|_{2}$ & $\ell_2$ norm\\
        $\Pr(\, \cdot \, )$ & Probability\\
        $K!$ & Factorial of positive integer $K$\\
        \bottomrule
	\end{tabular}\label{Table: Notations}
    \end{spacing}
\vspace{-10pt}
\end{table}

\section{Signal Model, MSE and CRB}\label{Signal Model}
Assuming $K$ far-field narrowband signals impinging on a linear array consisting of $M$ sensors located at $\{d_1,d_2,\cdots,d_{M}\}$ from directions $\bm{\theta} = [\theta_1,\theta_2,\cdots,\theta_K]^\mathrm{T}$, where, without loss of generality, the first $L$ ($1 \leq L \leq K$) sources are assumed to be mutually coherent, while the left $K-L$ sources are incoherent with each other and independent of the first $L$ sources.
The array observation data are modeled as
\begin{equation}\label{Eq:received signal vector}
\begin{aligned}
  \bm{x}(t) &= \sum_{k=1}^{L}\bm{a}(\theta_k) \beta_k s_1(t) + \sum_{k=L+1}^{K} \bm{a}(\theta_k) s_k(t) + \bm{n}(t)\\
            &= \bm{A}(\bm{\theta}) \bm{s}(t) + \bm{n}(t), \forall t = 1, 2, \cdots, T,
\end{aligned}
\end{equation}
where $\bm{A}(\bm{\theta}) = [\bm{a}(\theta_1),\bm{a}(\theta_2),\cdots,\bm{a}(\theta_K)]\in\mathbb{C}^{M\times K}$ is the steering matrix with the $k$-th column
\begin{equation}\label{Eq:steering vector}
  \bm{a}(\theta_k)=\left[e^{-j\frac{2\pi}{\lambda}d_1\sin\theta_k},\cdots,e^{-j\frac{2\pi}{\lambda}d_{M}\sin\theta_k}\right]^\mathrm{T}
\end{equation}
denoting the steering vector of the $k$-th source with $\lambda$ representing the wavelength,  $\bm{s}(t)\!=\![\bm{\beta}^{\mathrm{T}}s_{1}(t),s_{L+1}(t),\cdots,s_K(t)]^\mathrm{T} \! \in \! \mathbb{C}^{K}$ denotes the $K$ signals with $\bm{\beta} \!=\! [\beta_1,\cdots,\beta_L]^{\mathrm{T}} \in \mathbb{C}^{L}$
containing the coherent coefficient $\beta_{l}$ between the $l$-th ($1 \leq l \leq L$) coherent signal $\beta_l s_1(t)$ and the reference signal $s_1(t)$ (i.e., $\beta_1 = 1$),
and $\bm{n}(t) \sim \mathcal{CN}(\bm{0}_M,\sigma^2_n\bm{I}_{M})$ is the complex zero-mean additive white Gaussian noise independent of the signals with $\sigma^2_n$ denoting the noise power. Here, $T$ denotes the number of snapshots.

Without loss of generality, we make the following assumptions:

\vspace{-10pt}
\newtheorem{assumption}{Assumption}
\begin{assumption}\label{A1}
All the incoherent signals $s_k(t), k=1,L\!+\!1,\cdots,K$ are sampled from zero-mean, stationary complex Gaussian stochastic processes.
\end{assumption}
\vspace{-14pt}
\newtheorem{assumption2}{Assumption}
\begin{assumption}\label{A2}
The DOAs of the signals are distinct (say, $\theta_{\imath} \neq \theta_{\jmath}$ $\forall \imath \neq \jmath$).
\end{assumption}
\vspace{-14pt}
\newtheorem{assumption3}{Assumption}
\begin{assumption}\label{A3}
Each DOA follows a uniform distribution $\theta_k \sim \mathcal{U}[\vartheta_{\min},\vartheta_{\max}], \forall k = 1, 2,\cdots, K$.
\end{assumption}
\vspace{-6pt}

Thus, the theoretical covariance matrix of $\bm{x}(t)$ given $\bm{\theta}$ is
\begin{equation}\label{Eq:covariance matrix}
\begin{aligned}
   \bm{R}_{\bm{x}|\bm{\theta}}& =  \mathbb{E}   \left\{\bm{x}(t)\bm{x}^\mathrm{H} (t)   \right\}
                                = \bm{A}(\bm{\theta})\bm{\Sigma}\bm{A}^\mathrm{H}(\bm{\theta})+\sigma_n^2\bm{I}_{M},
\end{aligned}
\end{equation}
where
\begin{equation}\label{Eq:Sigma}
\begin{aligned}
  \bm{\Sigma} & = \mathbb{E}\left\{\bm{s}(t)\bm{s}^\mathrm{H} (t)\right\}
  = \mathrm{block \ diag} \left[\bm{\Sigma}_{co},\bm{\Sigma}_{in} \right]
\end{aligned}
\end{equation}
with
\begin{equation}\label{Eq:Sigma c}
\begin{aligned}
  \bm{\Sigma}_{co} = \bm{\beta}\bm{\beta}^{\mathrm{H}}\sigma_{1}^2
\end{aligned}
\end{equation}
denoting the covariance matrix corresponding to the first $L$ coherent sources, and
\begin{equation}\label{Eq:Sigma u}
\begin{aligned}
  \bm{\Sigma}_{in} = \mathrm{diag} \left[ \sigma_{L+1}^2,\cdots,\sigma_{K}^2 \right]
\end{aligned}
\end{equation}
denoting the power of the left $K-L$ incoherent sources.
The diagonal entries of matrix $\bm{\Sigma}$ denote the power of $K$ sources, where $|\beta_k|^2\sigma_1^2 \ (k=1, \cdots, L)$ denotes the power of the $k$-th coherent source, $\sigma_k^2$, while $\sigma_k^2 \ (k=L+1,\cdots,K)$ denotes the power of the $k$-th incoherent source.

Similar to other parameter estimation problems, the MSE defined as
\begin{equation}\label{Eq:MSE}
\begin{aligned}
  \text{MSE}  &  =   \frac{1}{K}  \sum_{k=1}^{K}  \mathbb{E}  \left\{  \left(\hat{\theta}_{k} - \theta_{k}\right)^{2}  \right\}
               =   \frac{1}{K} \mathrm{Tr}\{\bm{R}_{\bm{\epsilon}}\}
\end{aligned}
\end{equation}
is a well accepted metric to evaluate the estimation accuracy of DOA estimators, where $\hat{\theta}_{k}$ is the estimate of $\theta_k$, and
\begin{equation}\label{Eq:R epsilon}
\vspace{-1pt}
\begin{aligned}
 \bm{R}_{\bm{\epsilon}} & = \mathbb{E}\left\{ \bm{\epsilon} \bm{\epsilon}^{\mathrm{T}} \right\}
                         = \mathbb{E}\left\{ ( \hat{\bm{\theta}} - \bm{\theta} ) ( \hat{\bm{\theta}} - \bm{\theta} )^\mathrm{T} \right\}
\end{aligned}
\vspace{-2pt}
\end{equation}
is the error correlation matrix.
It is worth noting that, the definition of MSE for DOA estimation implies the following assumption:
\vspace{-6pt}
\newtheorem{assumption4}{Assumption}
\begin{assumption}\label{A4}
All DOAs make the same contribution to MSE calculation, namely, they have an equal weight.
\end{assumption}
\vspace{-6pt}

As a widely used lower bound in evaluating the variance of DOA estimators, CRB is the inverse of the Fisher information, where the $\imath\jmath$-th entry of the Fisher information matrix $\bm{J}$ w.r.t. $\bm{\theta}$ is given by \cite{van2002detection}
\begin{equation}\label{Eq:Fisher under}
\vspace{-1pt}
   \bm{J}_{\imath\jmath} = T \mathrm{Tr} \left\{  \frac{\partial \bm{R}_{\bm{x}|\bm{\theta}} }{\partial \theta_\imath} \bm{R}_{\bm{x}|\bm{\theta}}^{-1} \frac{\partial \bm{R}_{\bm{x}|\bm{\theta}} }{\partial \theta_\jmath} \bm{R}_{\bm{x}|\bm{\theta}}^{-1} \right\}.
   \vspace{-1pt}
\end{equation}
It is observed that, due to the uniform distribution, there is no \textit{a priori} information of $\bm{\theta}$ contributing to the Fisher information even if in Bayesian CRB (BCRB) \cite{van2002detection}, which makes CRB only tight in the asymptotic region.

\section{ZZB Preliminaries}\label{ZZB Preliminaries}
The extended ZZB for vector parameter estimation to lower bound the error correlation matrix is given in \cite[Eq. (32)]{Bell1997Extended}\footnote{The valley filling operation in \cite[Eq. (32)]{Bell1997Extended} is ignored since it is proved in \cite{Bell1996Explicit} that the valley filling operation does not have an effect in DOA estimation context.}
\begin{equation}\label{Eq:ZZB vector}
\begin{aligned}
 & \bm{w}^\mathrm{\!T}\! \bm{R}_{\bm{\epsilon}} \bm{w} \\
 & \!\! \geq  \frac{1}{2} \! \int_{0}^{\infty} \!\!\!\!\!  \max \limits_{ \bm{\delta} : \bm{w}^\mathrm{\!T} \bm{\delta} = h } \! \! \bigg[ \!\!  \int_{\mathbb{R}^{\!K}} \!\!\!\!   \left( f_{\bm{\theta}}(\bm{\varphi}) \! +\!  f_{\bm{\theta}} (\bm{\varphi} \!+\! \bm{\delta}) \right) \!
 P_{\mathrm{min}}(\bm{\varphi}, \bm{\varphi} \!+\! \bm{\delta}) d\bm{\varphi} \! \bigg]\! h dh, \\
\end{aligned}
\end{equation}
where $\bm{w} \in \mathbb{R}^{K}$ represents a normalized weight vector (i.e., $\|\bm{w}\|_{2} = 1$), $f_{\bm{\theta}}(\bm{\varphi}) \triangleq f(\bm{\theta})|_{\bm{\theta}=\bm{\varphi}}$ with $f(\bm{\theta})$ denoting the \textit{a priori} probability density function (PDF) of $\bm{\theta}$, and $P_{\mathrm{min}}(\bm{\varphi}, \bm{\varphi} + \bm{\delta})$ is the minimum probability of error of the binary hypothesis testing problem
\begin{equation}\label{detection problem}
\begin{aligned}
  &\mathcal{H}_0: \bm{\theta}=\bm{\varphi};  \qquad\quad \bm{x}\sim f( \bm{x} | \bm{\varphi}) \\
  &\mathcal{H}_1: \bm{\theta}=\bm{\varphi}+\bm{\delta}; \quad\;  \bm{x}\sim f( \bm{x} | \bm{\varphi} + \bm{\delta} )
\end{aligned}
\end{equation}
with
\begin{equation}\label{Pr of hypothesis 0}
\begin{aligned}
   \mathrm{Pr}(\mathcal{H}_0) &= \frac{f_{\bm{\theta}}(\bm{\varphi})} {f_{\bm{\theta}}(\bm{\varphi})+f_{\bm{\theta}}(\bm{\varphi}+\bm{\delta})}\\
   \mathrm{Pr}(\mathcal{H}_1) &= 1-\mathrm{Pr}(\mathcal{H}_0).
\end{aligned}
\end{equation}
Here, $f( \bm{x} | \bm{\varphi})$ and $f( \bm{x} | \bm{\varphi} + \bm{\delta} )$ are respectively the conditional PDFs of $\bm{x}$ given $\bm{\theta} = \bm{\varphi} $ and $\bm{\theta} = \bm{\varphi} + \bm{\delta} $. With Assumption \ref{A4}, the weight vector $\bm{w}$ in DOAs estimation becomes
\begin{equation}\label{Eq:equal weigh vector}
 \bm{w} = \frac{1}{\sqrt{K}}\bm{1}_{K}.
\end{equation}
Thus, \eqref{Eq:ZZB vector} can be written as
\begin{equation}\label{Eq:ZZB vector same weight}
\begin{aligned}
 \frac{1}{K}  \bm{1}_{K}^{\mathrm{T}} \bm{R}_{\bm{\epsilon}} \bm{1}_{K} \!
    \geq & \frac{1}{2} \! \int_{0}^{\infty} \!\!\!\! \max \limits_{ \bm{\delta} : \bm{1}_{\!K}^{\!\mathrm{T}}\bm{\delta} = \sqrt{K} h } \! \bigg[ \! \int_{\mathbb{R}^{\!K}} \!\!\!\! \left( f_{\bm{\theta}}(\bm{\varphi}) \!+\! f_{\bm{\theta}} (\bm{\varphi} \!+\! \bm{\delta}) \right) \!  \\
 & \quad\quad\quad\quad\quad\quad\   \!\times\! P_{\mathrm{min}}(\bm{\varphi}, \bm{\varphi} \!+\! \bm{\delta}) d\bm{\varphi} \bigg] h dh.
\end{aligned}
\end{equation}

For single source case (i.e., $K=1$), the vector $\bm{\theta}$ degrades to a scalar $\theta$, $\delta = h$, and \eqref{Eq:ZZB vector same weight} becomes
\begin{equation}\label{Eq:ZZB scalar}
\begin{aligned}
  & \mathbb{E}  \left\{   \left( \hat{\theta} - \theta  \right)^{2}  \right\}   \\
   & \geq  \!  \frac{1}{2} \!\!\int_{0}^{\infty} \! \! \! \! \int_{-\!\infty}^{+\!\infty} \! \!  \! \left( f_{\theta}(\varphi)  \!+ \! f_{\theta} (\varphi  \!+ \! h) \right)
 P_{\mathrm{min}}(\varphi, \varphi \!+\! h) d\varphi  h dh,
\end{aligned}
\end{equation}
which also has been investigated in \cite{Bell1996Explicit}. On the contrary, for multiple sources case (i.e., $K\geq2$), it's still challenging for ZZB to evaluate the performance of DOA estimators straightforwardly due to $K$-dimensional integration and optimization in \eqref{Eq:ZZB vector same weight}.

\begin{figure*}[!b]
\hrulefill
\vspace{2pt}
\centering
\setcounter{equation}{23}
\begin{equation}\label{PL}
\begin{aligned}
 P_{\mathrm{L}}
  \!=\! e^{\! T \left[ \ln \! \frac{ 4(1+M \|\bm{\beta}\|_2^2 \eta_1 ) }{ (2+M \|\bm{\beta}\|_2^2 \eta_1 )^{2} }
  + \left( \! \frac{M \|\bm{\beta}\|_2^2 \eta_1}{2+M \|\bm{\beta}\|_2^2 \eta_1 } \! \right)^{\!2}
  + \!\!\! \sum \limits_{k=L\!+\!1}^{K} \! \left[\ln \! \frac{4(1+M\eta_{k})}{(2+M\eta_{k})^2}
  + \left( \! \frac{M\eta_k}{2+M\eta_k} \! \right)^{\!2} \right] \right]}
  \mathcal{Q}\left( \! \sqrt{ 2T \! \left[ \left(\frac{M \|\bm{\beta}\|_2^2 \eta_1 }{2\!+\!M \|\bm{\beta}\|_2^2 \eta_1}\right)^{\!2} \!\!\! + \!\!\!\!\sum_{k=L\!+\!1}^{K} \!  \left(\frac{M\eta_k}{2\!+\!M\eta_k}\right)^{\!2}\right]} \right)
\end{aligned}
\end{equation}
\end{figure*}

\section{Derivation of ZZB for DOAs Estimation}\label{ZZB derivation}
In this section, we derive a ZZB for DOAs estimation in an explicit way, where the number of sources, the number of array sensors, the
number of snapshots, the \textit{a priori} distribution and SNRs of sources, array observation data, and the coherent coefficient are served as explicit factors. In subsection \ref{Generalizing}, we first follow the derivation framework in \cite{Bell1996Explicit} to generalize the ZZB for single source
DOA estimation to multiple sources DOA estimation, which, unfortunately, is invalid for lower bounding the MSE outside the asymptotic region. Considering the permutation ambiguity arising in MSE calculation for multiple sources DOA estimation, in subsection \ref{Modification}, we then introduce order statistics to make the ZZB global tight in evaluating the MSE for multiple sources DOA estimation.

\subsection{Generalized ZZB for Multiple Sources DOA Estimation}\label{Generalizing}
It is worth noting that, the utilization of Sherman-Morrison formula and matrix determinant lemma in \cite{Bell1996Explicit} makes it appropriate for single source case only, which, accordingly, hinders the ZZB derivation in multiple sources scenario. Instead, for the hybrid coherent/incoherent multiple sources considered in this paper, we introduce Woodbury matrix identity and Sylvester's determinant theorem to generalize the ZZB as an explicit function of multiple sources.

For the array observation data matrix $\bm{X} = [ \bm{x}(1) , \cdots, \bm{x}(T) ] \in \mathbb{C}^{M\times T}$, $P_{\mathrm{min}}(\bm{\varphi}, \bm{\varphi} + \bm{\delta})$ is lower bounded as \cite[p. 79]{van2001detection}
\setcounter{equation}{15}
\begin{equation}\label{Pmin}
\begin{aligned}
 &\!\!\!\!P_{\mathrm{min}} (\bm{\varphi}, \bm{\varphi} + \bm{\delta})  \\
 &\!\!\!\! \geq e^{ \left[ \mu(p;\bm{\delta}) +  \frac{1}{8} \frac {\partial^2 \mu(p;\bm{\delta})} {\partial p^2}    \right]}\mathcal{Q}\left( \frac{1}{2}\sqrt{\frac {\partial^2 \mu(p;\bm{\delta})} {\partial p^2}} \right) \Bigg|_{p=\frac{1}{2}}\\
 &\!\!\!\! \triangleq P(\bm{\delta}),
\end{aligned}
\vspace{2pt}
\end{equation}
where
\begin{equation}\label{Q}
 \mathcal{Q}(z) = \int_{z}^{\infty} \frac{1}{\sqrt{2\pi}} e^{-\frac{v^2}{2}} d v
\end{equation}
is the tail distribution function of the standard normal distribution, and
\begin{equation}\label{mu}
 \mu(p;\bm{\delta}) = \ln \int f( \bm{X} | \bm{\varphi} + \bm{\delta} ) ^ p f( \bm{X} | \bm{\varphi} ) ^ {1-p} d \bm{X}
\end{equation}
is the semi-invariant moment generating function.
According to Appendix \ref{AppA}, $\mu(p;\bm{\delta})|_{p=\frac{1}{2}}$ and $\frac {\partial^2 \mu(p;\bm{\delta})} {\partial p^2}\big|_{p=\frac{1}{2}}$ in \eqref{Pmin} are respectively approximated as
\begin{equation}\label{mu1/2 final}
\begin{aligned}
 \!\!   \mu(p;\bm{\delta})|_{p=\frac{1}{2}} \! \approx \!\! \left\{\!\!
    \begin{array}{lc}
        - \frac{1}{8} \bm{\delta}^\mathrm{T} \bm{J} \bm{\delta} , \! & \!\! \bm{\delta} \! \in \! \Delta \\
        T \! \left[ \ln \! \frac{ 4(1\!+\!M \|\bm{\beta}\|_2^2 \eta_1) }{ (2\!+\!M \|\bm{\beta}\|_2^2 \eta_1)^2 } \!+\!\!\!\!\!\sum \limits_{k=L\!+\!1}^{K}\!\!\!\! \ln \! \frac{4(1\!+\!M\eta_{k})}{(2\!+\!M\eta_{k})^2}\right] , \! & \!\!  \bm{\delta}\! \notin \! \Delta
    \end{array}
\!\!,\right.
\end{aligned}
\vspace{2pt}
\end{equation}
and
\begin{equation}\label{mu1/2 second derivative final}
\begin{aligned}
   \frac {\partial^2 \mu(p;\bm{\delta})} {\partial p^2}\bigg|_{p=\frac{1}{2}} \! \! \! \! \! \approx \!  \left\{\!\!
    \begin{array}{lc}
        \! \bm{\delta}^\mathrm{T} \bm{J} \bm{\delta} ,  & \!\! \bm{\delta} \! \in \!  \Delta  \\
        \! 8T \! \left[ \! \left( \! \frac{M \|\bm{\beta}\|_2^2 \eta_1}{2\!+\!M \|\bm{\beta}\|_2^2 \eta_1} \! \right)^{\!2} \!\!\! + \!\!\!\!\! \sum \limits_{k=L\!+\!1}^{K} \!\!\! \left( \! \frac{M\eta_k}{2\!+\!M\eta_k} \! \right)^{\!2} \! \right] \! , & \!\!  \bm{\delta} \! \notin \!  \Delta
    \end{array}
\!\!,\right.
\end{aligned}
\vspace{2pt}
\end{equation}
where $\eta_k = \frac{\sigma_{k}^2}{\sigma_n^2}$ denotes the SNR of the $k$-th source, and $\Delta$ denotes the region
\begin{equation}\label{small region}
 \Delta \!=\!  \left\{ \! \bm{\delta}\!: \bm{\delta}^\mathrm{T} \bm{J} \bm{\delta}  \! \leq \! 8T \! \left[ \! \left( \! \frac{M \|\bm{\beta}\|_2^2 \eta_1}{2\!+\!M \|\bm{\beta}\|_2^2 \eta_1} \right)^{\!\!2} \!\!\!+\! \!\!\!\sum_{k=L\!+\!1}^{K} \!\!\!\left(\frac{M\eta_k}{2\!+\!M\eta_k}\right)^{\!\!2} \right] \! \right\}\!.
\vspace{2pt}
\end{equation}

Accordingly, $P(\bm{\delta})$ in \eqref{Pmin} is approximated as
\setcounter{equation}{21}
\begin{equation}\label{Pmin final}
\begin{aligned}
 P(\bm{\delta}) \approx \left\{
    \begin{array}{lc}
        P_{\mathrm{S}}(\bm{\delta}),   & \bm{\delta} \in \Delta \\
        P_{\mathrm{L}},   & \bm{\delta} \notin \Delta
    \end{array}
\right.
\end{aligned}
\vspace{2pt}
\end{equation}
with
\begin{equation}\label{Ps}
\begin{aligned}
       P_{\mathrm{S}}(\bm{\delta}) = \mathcal{Q}\left( \frac{1}{2}\sqrt{\bm{\delta}^\mathrm{T} \bm{J} \bm{\delta}} \right),
\end{aligned}
\vspace{2pt}
\end{equation}
and $P_{\mathrm{L}}$ \eqref{PL} is shown at the bottom of this page.

It should be pointed out that, the $P_{\mathrm{S}}(\bm{\delta})$ derived here is appropriate for multiple sources case although it has the same expression as \cite[Eq. (D. 13)]{Bell1996Explicit}.
To be more specific, the determinants of $\mu(p;\bm{\delta})|_{p=\frac{1}{2}}$ in \cite{Bell1996Explicit} are calculated using matrix determinant lemma in \cite[Eq. (B. 16)]{Bell1996Explicit} and \cite[Eq. (B. 17)]{Bell1996Explicit}, from which the $P_{\mathrm{S}}(\bm{\delta})$ is finally derived. Considering the fact that the use of matrix determinant lemma in \cite{Bell1996Explicit} implies the single source assumption, the $P_{\mathrm{S}}(\bm{\delta})$ derived there is naturally appropriate for single source only.
On the contrary, we perform the Taylor expansion directly on $\mu(p;\bm{\delta})|_{p=\frac{1}{2}}$ and $\frac {\partial^2 \mu(p;\bm{\delta})} {\partial p^2}\big|_{p=\frac{1}{2}}$ that preserves the matrix determinants to derive the $P_{\mathrm{S}}(\bm{\delta})$, during which the assumption about number of sources has not been involved.
Therefore, the $P_{\mathrm{S}}(\bm{\delta})$ derived here is appropriate for both single source and multiple sources simultaneously.
Besides, we adopt Woodbury matrix identity and Sylvester's determinant theorem rather than Sherman-Morrison formula and matrix determinant lemma in \cite{Bell1996Explicit} to derive the $P_{\mathrm{L}}$, which, not only makes it work for multiple sources, but also enables us for the first time to formulate it as an explicit function of coherent coefficients $\bm{\beta}$.
Moreover, the derived $P_{\mathrm{L}}$ is also a function of the number of sources, which is approximately\footnote{Our derivation of $P_{\mathrm{L}}$ is consistent with \cite[Eq. (C. 13)]{Bell1996Explicit}. However, an extra approximation was made in \cite[Eq. (C. 13)]{Bell1996Explicit}. Hence, we mention ``approximately" here.} same as that in \cite{Bell1996Explicit} when $K=1$.

\begin{figure*}[!b]
\vspace{-8pt}
\hrulefill
\vspace{1pt}
\centering
\setcounter{equation}{30}
\begin{equation}\label{Eq:hs approx}
\begin{aligned}
\tilde{h} \approx  \min  \left[   \sqrt{ \frac{8T \bm{1}_{K}^{\mathrm{T}} \bm{J}^{-1} \bm{1}_{K}  }{K} \left[ \left( \frac{M \|\bm{\beta}\|_2^2 \eta_1}{2+M \|\bm{\beta}\|_2^2 \eta_1}\right)^2+ \sum_{k=L+ 1}^{K}   \left(\frac{M\eta_{k}} {2 +  M\eta_{k}} \right)^{  2} \right] } ,   \sqrt{K}\zeta \right].
\end{aligned}
\end{equation}
\end{figure*}

Since $K$ DOAs are independent, the $K$-dimensional integration in \eqref{Eq:ZZB vector same weight}  becomes
\setcounter{equation}{24}
\begin{equation}\label{K dimensional integration}
\begin{aligned}
& \frac{1}{2} \int_{\mathbb{R}^K}  \left( f_{\bm{\theta}}(\bm{\varphi}) +  f_{\bm{\theta}} (\bm{\varphi} + \bm{\delta}) \right)  P_{\mathrm{min}}(\bm{\varphi}, \bm{\varphi} + \bm{\delta}) d\bm{\varphi} \\
& \geq \frac{P(\bm{\delta})}{\zeta^K} \int_{\Phi}  d\bm{\varphi}  \\
&  =  \frac{P(\bm{\delta})}{\zeta^K} \prod\limits_{k=1}^{K} (\zeta-|\delta_k|),
\end{aligned}
\vspace{1pt}
\end{equation}
where
\begin{equation}\label{Eq:Phi region}
\begin{aligned}
 \Phi = \left\{ \bm{\varphi} | \varphi_k \in \left[\vartheta_{\mathrm{min}},\vartheta_{\mathrm{max}} - |\delta_k|\right], k = 1,2, \cdots ,K \right\}
\end{aligned}
\vspace{1pt}
\end{equation}
is the $K$-dimensional integration region,  $\zeta = \vartheta_{\max}-\vartheta_{\min}$ denotes the range of DOAs, and $|\delta_k|$ is the absolute value of the $k$-th element in $\bm{\delta}$.

According to Appendix \ref{AppB}, the generalized ZZB for multiple sources DOA estimation is derived as
\begin{equation}\label{ZZB final}
\vspace{2pt}
\begin{aligned}
\!\!\!\!\frac{1}{K}  \bm{1}_{K}^\mathrm{T} \bm{R}_{\bm{\epsilon}} \bm{1}_{K} \! \geq \! \frac{12 P_{\mathrm{L}} \bm{1}_{K}^{\mathrm{T}} \bm{R}_{\bm{\theta}}  \bm{1}_{K} }{(K +\!1)(K\!+\!2)}  \!+\! \Gamma_{\frac{3}{2}} \! \left( \tilde{u} \right) \frac{\bm{1}_{K}^{\mathrm{T}} \bm{J}^{-1}  \bm{1}_{K}}{K}   ,
\end{aligned}
\vspace{2pt}
\end{equation}
where
\begin{equation}\label{R theta}
\begin{aligned}
\bm{R}_{\bm{\theta}} = \frac{\zeta^2}{12}\bm{I}_{K}
\end{aligned}
\vspace{2pt}
\end{equation}
is the \textit{a priori} covariance matrix of $\bm{\theta}$,
\begin{equation}\label{incomplete Gamma function}
\vspace{2pt}
\begin{aligned}
\Gamma_{\frac{3}{2}}( \tilde{u}) = \frac{1}{\Gamma\left(\frac{3}{2}\right)} \int_{0}^{ \tilde{u}} e^{-\xi} \xi^{\frac{1}{2}} d \xi
\end{aligned}
\end{equation}
is the normalized incomplete Gamma function with $\Gamma(\frac{3}{2}) = \frac{\sqrt{\pi}}{2}$ and
\begin{equation}\label{us}
\begin{aligned}
\tilde{u} = \frac{K\tilde{h}^2}{8\bm{1}_{K}^{\mathrm{T}} \bm{J}^{-1} \bm{1}_{K}}.
\end{aligned}
\vspace{-2pt}
\end{equation}
Here, $\tilde{h}$ \eqref{Eq:hs approx} shown at the bottom of this page.
Obviously, the generalized ZZB in \eqref{ZZB final} lower bounds the error correlation matrix, one special case of which is the MSE we sought. Specifically, considering that the MSE defined in \eqref{Eq:MSE} is only related to the diagonal elements of the error correlation matrix $\bm{R}_{\bm{\epsilon}}$ (also, the diagonal elements in $\bm{J}^{-1}$ \cite{Kay1993FundamentalsVolI}), the MSE of DOAs estimation can be lower bounded by the generalized ZZB as
\setcounter{equation}{31}
\begin{eqnarray}\label{ZZB MSE final}
\text{MSE}
 \geq  \frac{12 P_{\mathrm{L}}\mathrm{Tr}\{\bm{R}_{\bm{\theta}}\}}{(K + 1)(K + 2)}  + \Gamma_{\frac{3}{2}}   \left( \tilde{u} \right) \frac{\mathrm{Tr}\{ \bm{J}^{-1} \}}{K}  .
\end{eqnarray}
Obviously, only the first term of the generalized ZZB is a function of the \textit{a priori} distribution of $\bm{\theta}$ via $\bm{R}_{\bm{\theta}}$, while the second term is a function of the Fisher information matrix $\bm{J}$.

\subsection{ZZB for Estimation with Permutation Ambiguity}\label{Modification}
In subsection \ref{Generalizing}, we generalized the ZZB of single source DOA estimation to multiple sources DOA estimation, which, however, is invalid in the \textit{a priori} performance region. The reason is that, the ordering process is required by multiple sources DOA estimators for the elimination of permutation ambiguity, which implicitly changes the \textit{a priori} distribution of DOAs. To better illustrate the effect of the permutation ambiguity on MSE calculation, we consider the following example.

\textit{Example 1}:
Assuming there are two signals impinging on the array from $30^\circ$ and $45^\circ$, while their estimates are respectively $29^\circ$ and $44^\circ$. Accordingly, the root MSE (RMSE) is calculated as $\sqrt{\frac{ (30^\circ - 29^\circ)^2  +  (45^\circ - 44^\circ)^2 }{2}}  =  1.00^\circ$. However, when permutation ambiguity occurs, the outputs of DOAs estimator become $44^\circ$ and $29^\circ$, leading to a wrong RMSE of $\sqrt{\frac{ (30^\circ  -  44^\circ)^2 +  (45^\circ  -  29^\circ)^2 }{2}}  \approx  15.03^\circ$.

Although permutation ambiguity is ubiquitous for DOAs estimator, its impact on the MSE calculation is more prominent in the asymptotic region. Generally speaking, in the asymptotic region, the estimates of DOAs are around the true values, which leads to a small estimation error. On the contrary, the farther away from the asymptotic region, the larger the estimation error is. Obviously, to obtain the correct MSE, the outputs of DOAs estimators $\hat{\bm{\theta}}$ should not be directly matched to the true DOAs $\bm{\theta}$ during MSE calculation.
To eliminate the permutation ambiguity, both the estimated DOAs and the true DOAs are respectively sorted in ascending order before calculating the MSE. As such, the MSE should be calculated as
\begin{equation}\label{MSE for spectral}
  \text{MSE} = \frac{1}{K} \sum_{k=1}^{K} \mathbb{E}\left\{ \left(\hat{\theta}_{(k)} - \theta_{(k)}\right)^2 \right\} ,
\vspace{-3pt}
\end{equation}
where $\hat{\theta}_{(k)}$ and $\theta_{(k)}$ are the $k$-th smallest order statistic in $\hat{\bm{\theta}}$ and $\bm{\theta}$, respectively. Here, the subscript $(k)$ enclosed in parentheses indicates the $k$-th order statistic of the sample.

Although the ordering process enables an accurate MSE calculation in the asymptotic region, it simultaneously has an effect on the MSE outside the asymptotic region because it implicitly changes the \textit{a priori} distribution of DOAs $\bm{\theta}$. Actually, different DOAs are naturally assumed to be independent with each other, while such an ordering process introduces an extra dependence among the DOAs.
For example, in the \textit{a priori} performance region,
each $\theta_k$ tends to be randomly estimated according to its \textit{a priori} distribution $\theta_k \sim \mathcal{U}[\vartheta_{\min},\vartheta_{\max}]$, and the MSE of $\theta_k$ converges to its \textit{a priori} variance $\sigma_{\theta_{k}}^2$.
After the ordering process, $\theta_{(k)}$ tends to be randomly estimated in the interval $[\theta_{(k-1)},\theta_{(k+1)}]$\footnote{The interval becomes $[\vartheta_{\mathrm{min}},\theta_{(2)}]$ and $[\theta_{(K-1)},\vartheta_{\mathrm{max}}]$ for $\theta_{(1)}$ and $\theta_{(K)}$, respectively.} rather than original $[\vartheta_{\min},\vartheta_{\max}]$, and the MSE of $\theta_{(k)}$ converges to its \textit{a priori} variance $\sigma_{ \theta_{(k)} }^2$.
Obviously, with the number of sources $K$ increasing, the average range of $\theta_{(k)}$ becomes narrower compared with that of the original $\theta_{k}$, making the MSE scale down in the \textit{a priori} performance region due to the ordering process.

According to aforementioned discussion, the effect of ordering process on the MSE calculation is formulated as the ratio
\begin{equation}\label{modification factor}
\vspace{-1pt}
\begin{aligned}
  \kappa = \frac{\mathrm{Tr}\{\bm{R}_{(\bm{\theta})}\}}{\mathrm{Tr}\{\bm{R}_{\bm{\theta}}\}}
\end{aligned}
\vspace{-1pt}
\end{equation}
between the trace of the $\textit{a priori}$ covariance matrix of the order statistics $\{\theta_{(1)},\cdots,\theta_{(K)} \}$, i.e.,
\begin{equation}\label{R theta order}
\vspace{-1pt}
\begin{aligned}
 \mathrm{Tr}\{\bm{R}_{(\bm{\theta})}\} = \sum_{k=1}^{K}  \sigma_{ \theta_{(k)} }^2,
\end{aligned}
\vspace{-1pt}
\end{equation}
and that of the $\textit{a priori}$ covariance matrix of the original $\bm{\theta}$ before the ordering process, say, $\mathrm{Tr}\{\bm{R}_{\bm{\theta}}\}$.

According to the \textit{a priori} PDF of $\theta_{(k)}$ \cite[p.229]{Casella2002statistical}
\begin{equation}\label{pdf order}
\vspace{-1pt}
\begin{aligned}
   f(\theta_{(k)}) = \frac{K! \left( \frac{\theta_{(k)}}{\zeta} \right)^{k-1} \left( 1 - \frac{\theta_{(k)}}{\zeta} \right)^{K-k} }{\zeta(k-1)!(K-k)!},
\end{aligned}
\vspace{-1pt}
\end{equation}
the \textit{a priori} variance of $\theta_{(k)}$ is
\begin{equation}\label{var}
\vspace{-1pt}
\begin{aligned}
   \sigma_{\theta_{(k)} }^2  =  &  \int_{0}^{\zeta}   \left( \theta_{(k)} - \mathbb{E} \left\{ \theta_{(k)}\right\}\right)^2 f(\theta_{(k)}) d \theta_{(k)} \\
   =& \ \frac{\zeta^2(K+1-k)k}{(K+1)^2(K+2)},
\end{aligned}
\vspace{-1pt}
\end{equation}
where
\begin{equation}\label{mean}
\vspace{-1pt}
\begin{aligned}
   \mathbb{E} \left\{ \theta_{(k)}\right\} =& \int_{0}^{\zeta} \theta_{(k)} f(\theta_{(k)}) d \theta_{(k)} = \frac{ k \zeta}{K+1}.
\end{aligned}
\vspace{-1pt}
\end{equation}
Thus, \eqref{R theta order} becomes
\begin{equation}
  \mathrm{Tr}\{\bm{R}_{(\bm{\theta})}\} = \frac{K\zeta^2}{6(K+1)},
\end{equation}
and correspondingly, the ratio $\kappa$ in \eqref{modification factor} becomes
\begin{equation}\label{modification factor 2}
\begin{aligned}
  \kappa = \frac{2}{K+1},
\end{aligned}
\end{equation}
which is a function of the number of sources $K$. For single source case (i.e., $K=1$), we have $\kappa = 1$, namely, there is no permutation ambiguity. Furthermore, the ratio $\kappa$ gets smaller with the number of sources increasing, indicating that the MSE scales down more in the \textit{a priori} performance region after the ordering process. The reason is that, the average interval $[\theta_{(k-1)},\theta_{(k+1)}]$ gets narrower with the number of sources increasing.

In order to make the ZZB valid in the \textit{a priori} performance region for evaluating multiple sources DOA estimation, we incorporate the ratio $\kappa$ \eqref{modification factor 2} into the first term of \eqref{ZZB MSE final} to obtain the ZZB expression as
\begin{eqnarray}\label{ZZB final2}
   \text{MSE}  \geq  2 P_{\mathrm{L}} \frac{ K\zeta^2}{(K + 1)^2(K + 2)}   + \Gamma_{\frac{3}{2}}   \left( \tilde{u} \right)\frac{\mathrm{Tr}\{\bm{J}^{-1}\}}{K} .
\end{eqnarray}
Obviously, same as the generalized ZZB \eqref{ZZB MSE final}, the derived ZZB is also an explicit function of the number of sources, and depends on the number of snapshots, the number of array sensors, the \textit{a priori} distribution and SNRs of sources, array observation data via the Fisher information matrix, and the coherent coefficient $\bm{\beta}$.

In the \textit{a priori} performance region, the ZZB converges to the \textit{a priori} bound (APB) as
\begin{equation}\label{ZZB converge low}
\begin{aligned}
   \lim \limits_{\forall \eta_{k} \to 0} \text{MSE}   \geq \frac{K\zeta^2}{(K + 1)^2(K + 2)},
\end{aligned}
\end{equation}
since
\begin{equation}\label{ZZB converge low2}
\begin{aligned}
   \lim \limits_{\forall \eta_{k} \to 0} 2 P_{\mathrm{L}} = 1,
\end{aligned}
\end{equation}
and
\begin{equation}\label{ZZB converge low4}
\begin{aligned}
   \lim \limits_{\forall \eta_{k} \to 0} \Gamma_{\frac{3}{2}}\left( \tilde{u} \right) = 0.
\end{aligned}
\end{equation}
Obviously, the estimation accuracy of DOA estimator cannot be further improved by simply increasing the number of array sensors and/or the number of snapshots when the SNR tends to zero.

In the asymptotic region, the ZZB converges to the CRB as
\begin{equation}\label{ZZB converge high}
\begin{aligned}
   \lim \limits_{\forall \eta_{k} \to +\infty}   \text{MSE} \geq \frac{\mathrm{Tr}\{\bm{J}^{-1}\}}{K} ,
\end{aligned}
\end{equation}
since
\begin{equation}\label{ZZB converge high2}
\begin{aligned}
   \lim \limits_{\forall \eta_{k} \to +\infty} 2 P_{\mathrm{L}} = 0
\end{aligned}
\end{equation}
and
\begin{equation}\label{ZZB converge high3}
\begin{aligned}
   \lim \limits_{\forall \eta_{k} \to +\infty} \Gamma_{\frac{3}{2}}\left( \tilde{u} \right) = 1.
\end{aligned}
\end{equation}
Now it is clear that, the ZZB in \eqref{ZZB final2} is a linear combination between the APB and the CRB, whose coefficients are $2 P_{\mathrm{L}}$ and $\Gamma_{\frac{3}{2}} \! \left( \tilde{u} \right)$, respectively.

\begin{figure*}[t!]
\begin{minipage}[t]{0.49\linewidth}
\centering
\includegraphics[width=0.99\textwidth]{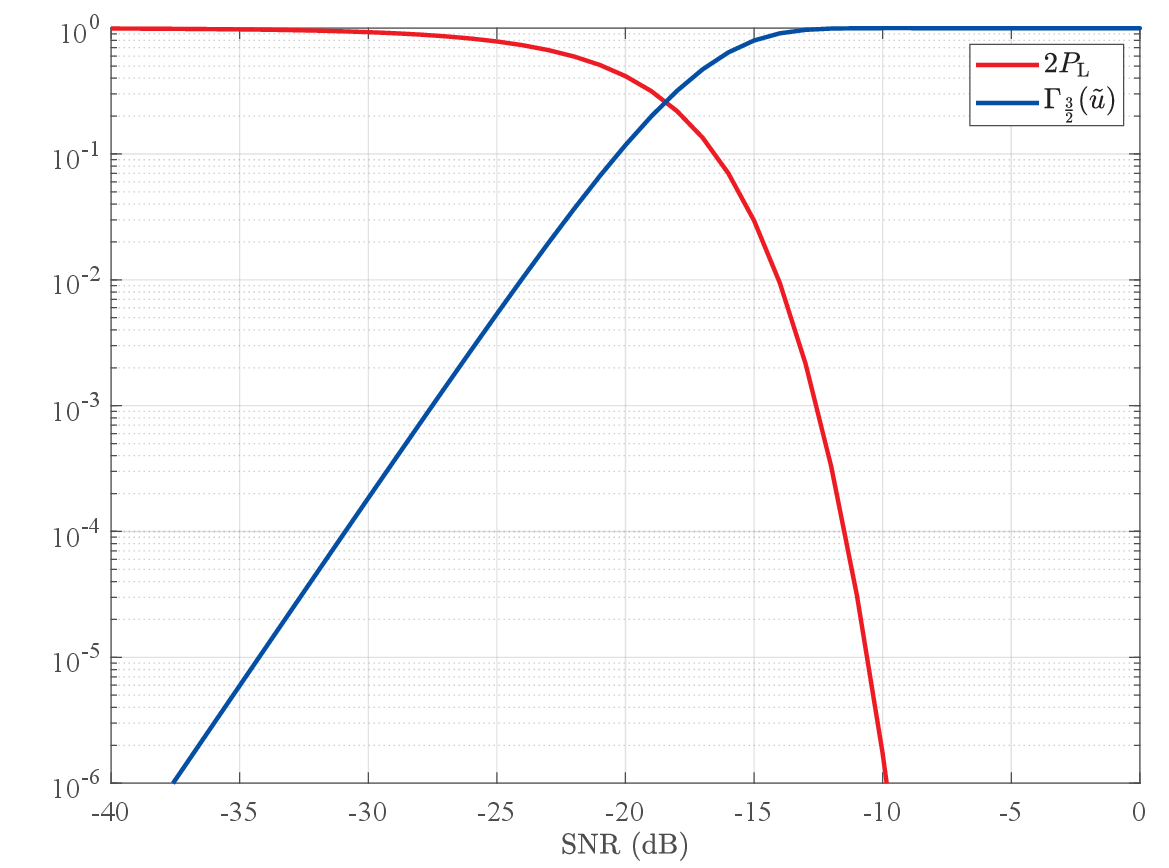}
\caption{Combination coefficients versus the SNR for single source.}\label{Fig:w}
\end{minipage}
\hfill
\begin{minipage}[t]{0.49\linewidth}
\centering
\includegraphics[width=0.99\textwidth]{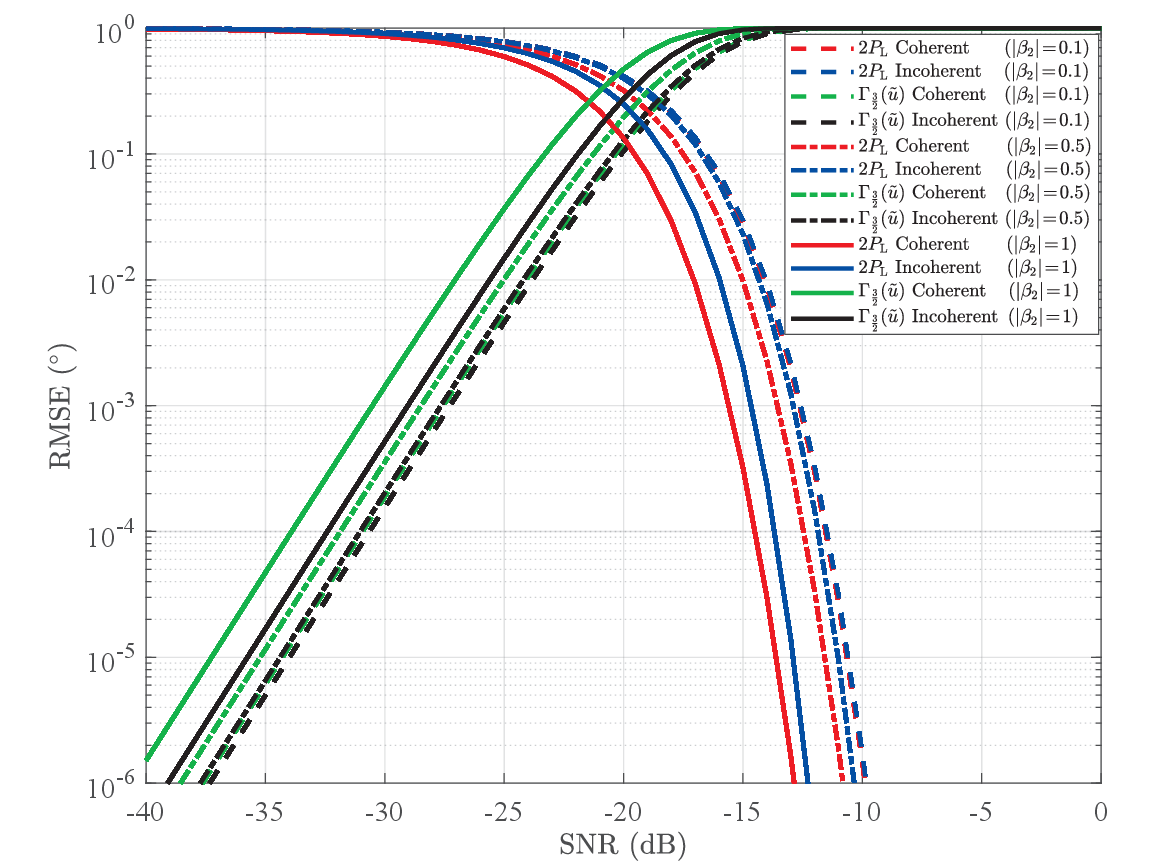}
\caption{Combination coefficients versus the SNR for two sources with different coherent coefficients.}\label{Fig:w coherent}
\end{minipage}
\end{figure*}

For fully incoherent sources (say, $L = 1$), $P_{\mathrm{L}}$ \eqref{PL} and $\tilde{u}$ \eqref{us} respectively degrade to
\begin{equation}\label{PL incoherent}
\begin{aligned}
 P_{\mathrm{L}}
  \!=\! e^{\! T \!\! \sum \limits_{k=1}^{K} \! \left[\ln \! \frac{4(1+M\eta_{k})}{(2+M\eta_{k})^2}
  + \left( \! \frac{M\eta_k}{2+M\eta_k} \! \right)^{\!2} \right] }
  \mathcal{Q}\!\left( \! \! \sqrt{ 2T \! \sum_{k=1}^{K} \!  \left(\frac{M\eta_k}{2\!+\!M\eta_k}\right)^{\!\!2}} \right)\!,
\end{aligned}
\end{equation}
and
\begin{equation}\label{Eq:us approx incoherent}
\begin{aligned}
\tilde{u}  \approx \min   \left[  T \sum_{k=1}^{K}   \left( \frac{M\eta_{k}} {2  +  M\eta_{k}} \right)^{  2}  ,  \frac{K^2\zeta^2}{8\bm{1}_{K}^{\mathrm{T}} \bm{J}^{ - 1} \bm{1}_{K}} \right] .
\end{aligned}
\end{equation}
When $K$ sources have the same SNR $\eta$, $P_{\mathrm{L}}$ and $\tilde{u}$ further become
\begin{equation}\label{PL equal SNR}
\begin{aligned}
\!\! P_{\mathrm{L}}  =   e^{ KT \left[ \ln  \frac{4(1+M\eta)} {(2+M\eta)^{2}} +  \left( \frac{M\eta} {2 + M\eta} \right)^{ 2} \right] } \mathcal{Q} \left( \sqrt{2KT} \frac{ M\eta} {2 + M\eta}  \right) ,
\end{aligned}
\end{equation}
and
\begin{equation}\label{us equal SNR}
\begin{aligned}
\tilde{u} & \approx \min   \left[ KT \left( \frac{M\eta} {2  +  M\eta} \right)^{  2} , \frac{K^2\zeta^2}{8\bm{1}_{K}^{\mathrm{T}} \bm{J}^{-1} \bm{1}_{K}} \right].
\end{aligned}
\end{equation}
Furthermore, when $K=1$, the ZZB is simplified to
\begin{equation}\label{ZZB single source}
\begin{aligned}
   & \mathbb{E}  \left\{  \left(\hat{\theta} -   \theta  \right)^{2}   \right\}   \\
   &   \geq \! 2  e^{ \! T \! \left[ \ln \! \frac{4(1+M\eta)} {(2+M\eta)^2} +  \left( \! \frac{M\eta} {2 + M\eta} \! \right)^{\!2} \right] } \mathcal{Q}\! \left( \! \sqrt{2T} \frac{ M\eta} {2 \!+\! M\eta} \! \right)  \sigma_{\!\theta}^{2}  + \Gamma_{\frac{3}{2}} \left( \tilde{u} \right) J^{-1} \!,
\end{aligned}
\end{equation}
where $\sigma_\theta^2=\frac{\zeta^2}{12}$ denotes the \textit{a priori} variance of $\theta$, and $J$ is the scalar Fisher information. It is worth pointing out that, the ZZB \eqref{ZZB single source} is consistent with the one \cite[Eq. (C.30)]{Bell1996Explicit}, if we ignore the implicit condition $\tilde{h}<\sqrt{K}\zeta$ such that
\begin{equation}\label{us single source}
\begin{aligned}
\tilde{u}   \approx  T \left(\frac{M\eta} {2  +   M\eta} \right)^{  2},
\end{aligned}
\end{equation}
and introduce the approximation \cite[Eq. (C.14)]{Bell1996Explicit}.

In Fig. \ref{Fig:w}, we compare combination coefficients of the ZZB versus the SNR, where a single source is assumed to follow \textit{a priori} distribution $\mathcal{U}[-90^{\circ},90^{\circ}]$ w.r.t. a uniform linear array (ULA) consisting of $M=20$ sensors. The number of snapshots is fixed to $T=40$, and $10,000$ Monte-Carlo trials are performed for each SNR point. It is observed that the coefficient $2 P_{\mathrm{L}}$ decreases from $1$ in the \textit{a priori} performance region to $0$ in the asymptotic region as analyzed, while the coefficient $\Gamma_{\frac{3}{2}} \! \left( \tilde{u} \right)$ has an opposite trend. That is to say, the APB dominates the ZZB in the \textit{a priori} performance region, but sharply vanishes in the asymptotic region. On the contrary, the CRB is ignorable in the \textit{a priori} performance region, but gradually dominates the ZZB from the transition region to the asymptotic region.

Unlike in the single source case, both combination coefficients $2 P_{\mathrm{L}}$ and $\Gamma_{\frac{3}{2}} \! \left( \tilde{u} \right)$ also depend on the coherent coefficient $\bm{\beta}$ under coherent multiple sources case. Hence, in Fig. \ref{Fig:w coherent}, we compare both combination coefficients versus the SNR for two coherent sources, where different magnitudes of the coherent coefficient $\beta_2$ are considered. Meanwhile, the corresponding combination coefficients for incoherent sources are also plotted as a reference, where the second source with power $\sigma_2^2 = |\beta_2|^2\sigma_1^2$ is independent of the first one. It is observed that, the difference between coherent and incoherent combination coefficients becomes larger with $|\beta_2|$ increasing. The reason is that, larger $|\beta_2|$ leads to larger norm $\|\bm{\beta}\|_2$ in \eqref{PL} and \eqref{Eq:hs approx}, which makes the combination coefficients $2 P_{\mathrm{L}}$ and $\Gamma_{\frac{3}{2}} \! \left( \tilde{u} \right)$ in coherent case more distinguishable from those in incoherent case.
On the other hand, with $|\beta_2|$ increasing, $2 P_{\mathrm{L}}$ and $\Gamma_{\frac{3}{2}} \! \left( \tilde{u} \right)$ respectively decreases from $1$ and reaches to $1$ at lower SNR,
indicating that the ZZB will leave the \textit{a priori} performance region to enter the asymptotic region at lower SNR. Furthermore, for the same $|\beta_2|$, $2 P_{\mathrm{L}}$ and $\Gamma_{\frac{3}{2}} \! \left( \tilde{u} \right)$ in coherent case respectively
decreases from $1$ and reaches to $1$ earlier than those in incoherent case, indicating that the coherent ZZB will leave from the APB to touch the CRB earlier than in incoherent case.

\textit{Remark}:
For underdetermined DOA estimation (say, $K>M$) using virtual coarray, the $\imath \jmath$-th entry of the coarray Fisher information matrix w.r.t. $\bm{\theta}$ is given by \cite[Eq. 37]{Liu2017Cramer},
\begin{equation}\label{Coarray FIM}
\vspace{-3pt}
\begin{aligned}
\tilde{\bm{J}}_{\imath\jmath}
  \!= \! \displaystyle{ T \!  \left[ \text{vec} \!  \left( \! \frac{ \partial \bm{R}_{\bm{x}|\bm{\theta}} }{ \partial \theta_{\imath} } \!\! \right)   \! \right]^{\mathrm{\!H}}\!\!   \left( \bm{R}_{\bm{x}|\bm{\theta}}^{\mathrm{T}} \! \otimes\!  \bm{R}_{\bm{x}|\bm{\theta}} \right) ^{\! -\!1} \!\!
  \text{vec}\!  \left( \! \frac{ \partial \bm{R}_{\bm{x}|\bm{\theta}} }{ \partial \theta_{\jmath} } \!\! \right) }.
\end{aligned}
\vspace{-1pt}
\end{equation}
to avoid the rank defect problem, where $\text{vec}(\, \cdot \,)$ denotes the vectorization operator, and $\otimes$ denotes the Kronecker product. By substituting the coarray Fisher information matrix into \eqref{ZZB final2}, the corresponding coarray ZZB can be used to evaluate underdetermined DOA estimation in a wide range of SNR from the \textit{a priori} performance region to the asymptotic region.

\vspace{-2pt}
\section{Simulations}\label{Simulations}
In this section, we evaluate the ZZB for multiple sources DOA estimation, where both the CRB and the APB are also plotted for reference.
Since ZZB requires the \textit{a priori} distribution, we assume each DOA to follow a uniform distribution $\mathcal{U}[-60^{\circ}, 60^{\circ}]$ in our simulations.
We run $\mathcal{L} = 10,000$ Monte-Carlo trials for each SNR point to obtain the involved bounds and the RMSE
\begin{equation}\label{RMSE}
\vspace{-2pt}
\begin{aligned}
   \text{RMSE} = \sqrt{\frac{1}{\mathcal{L}K}\sum_{\ell=1}^{\mathcal{L}}\sum_{k=1}^{K}\left(\hat{\theta}_{\ell,(k)}-\theta_{\ell,(k)}\right)^2}
\end{aligned}
\vspace{-2pt}
\end{equation}
of DOAs estimation. Here, $\hat{\theta}_{\ell,(k)}$ is the estimate of $\theta_{\ell,(k)}$, the $k$-th DOA after ordering process in the $\ell$-th Monte-Carlo trial.
To keep sources resolvable, we assume a minimum interval of $10^{\circ}$ in each $\bm{\theta}$ in multiple sources case, i.e., $K$ DOAs in each trial are sampled with at least $10^{\circ}$ separation. Unless otherwise specified, the number of snapshots is fixed to $T=40$, and a ULA with $M=20$ half-wavelength inter-element spacing sensors is adopted in our simulations, although any linear array configuration is applicable.

\begin{figure*}[t!]
\begin{minipage}[t]{0.49\linewidth}
\centering
\includegraphics[width=0.99\textwidth]{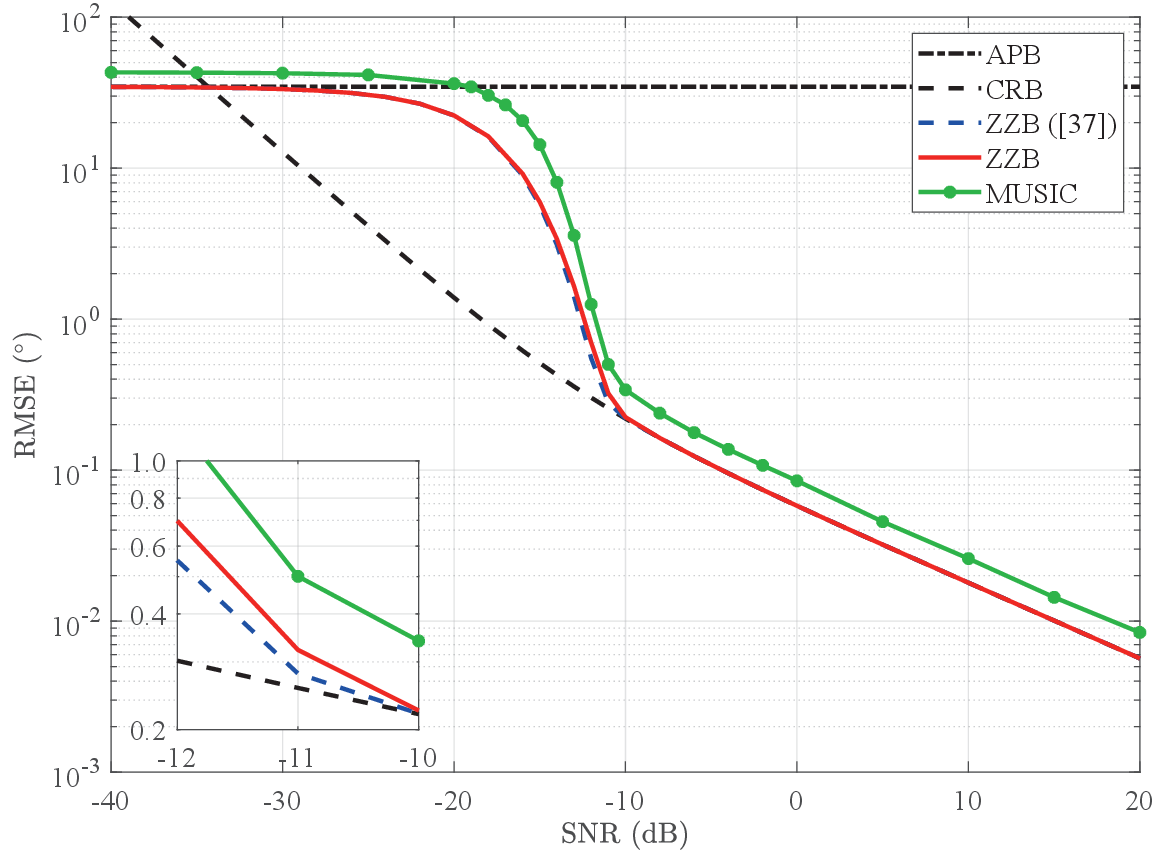}
\caption{Comparison of bounds for single source case ($K = 1$).}\label{Fig:K1}
\end{minipage}
\hfill
\begin{minipage}[t]{0.49\linewidth}
\centering
\includegraphics[width=0.99\textwidth]{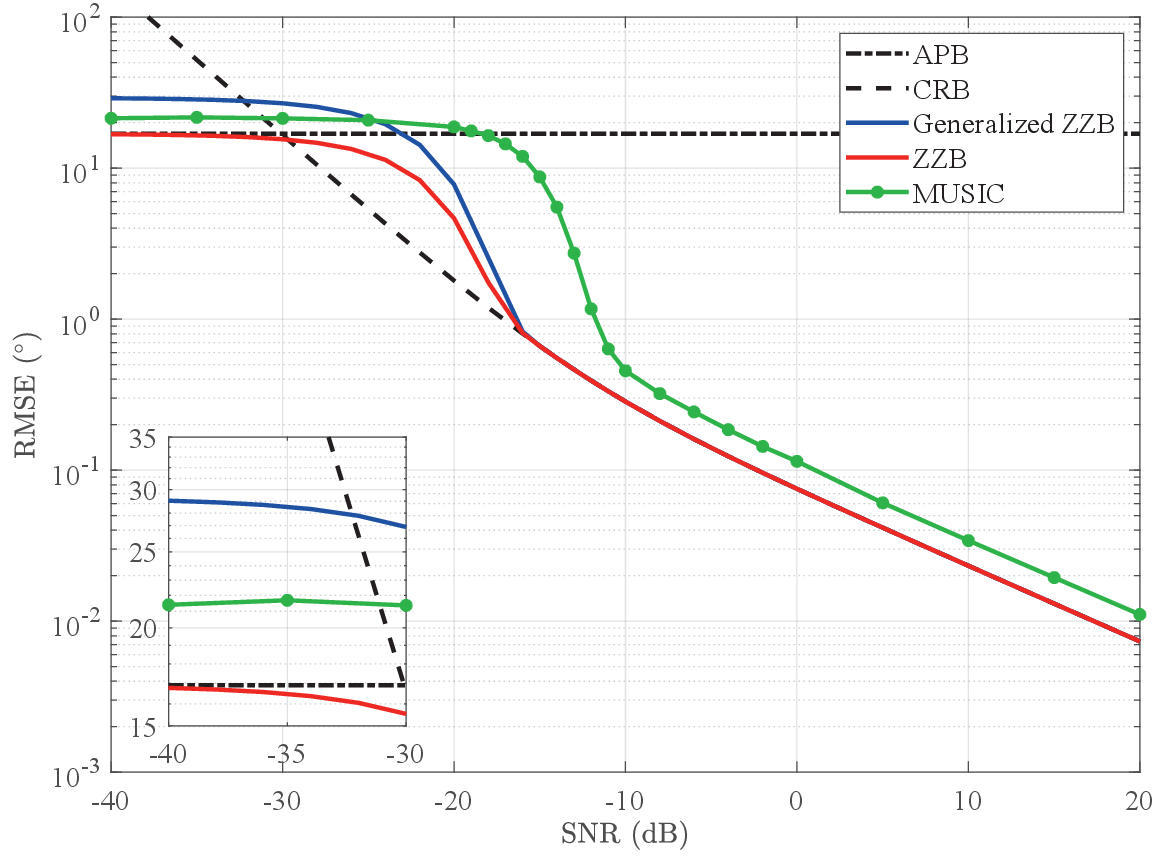}
\caption{Comparison of bounds for multiple sources case ($K = 5$).}\label{Fig:K5}
\end{minipage}
\end{figure*}

Considering that the ZZB in \cite{Bell1996Explicit} is only appropriate for the single source case, we first compare the derived ZZB \eqref{ZZB single source} with the ZZB in \cite{Bell1996Explicit} under the single source assumption. Meanwhile, the MUSIC algorithm is also plotted for a reference.
It is observed from Fig. \ref{Fig:K1} that, due to the extra approximation \cite[Eq. (C.14)]{Bell1996Explicit}, the explicit ZZB in \cite{Bell1996Explicit} enters the asymptotic region at a slightly lower input SNR than the derived ZZB, which is consistent with the numerical integration of \cite[Fig. 4]{Bell1996Explicit}. Besides, both the derived ZZB and the ZZB in \cite{Bell1996Explicit} converge to the APB in the \textit{a priori} performance region and the CRB in the asymptotic region, respectively.
Furthermore, the ZZB provides a tighter bound than the CRB outside the asymptotic region, and also predicts the threshold entering the asymptotic region for DOA estimation algorithms (e.g., MUSIC). On the contrary, since no \textit{a priori} information is used, the CRB provides a loose bound outside the asymptotic region, and finally divergence exceeds the MSE of the MUSIC algorithm in the \textit{a priori} performance region.

Then, we evaluate the derived ZZB under fully incoherent multiple sources assumption, where all $K$ sources are assumed to have the same SNRs.
It is observed from Fig. \ref{Fig:K5} that, both the generalized ZZB and the ZZB enter the asymptotic region at the same SNR, then keep consistent with the CRB in the asymptotic region. However, below the asymptotic region threshold, the generalized ZZB increases faster than the ZZB before entering the \textit{a priori} performance region, and finally convergence exceeds the MSE of the MUSIC algorithm. The CRB still provides a loose bound in the transition region, and finally divergence exceeds the MSE of the MUSIC algorithm. On the contrary, benefited from the ratio $\kappa$ \eqref{modification factor 2} considering the ordering process required in multiple sources DOA estimation, the ZZB is consistently lower than the generalized ZZB outside the asymptotic region, and keeps valid to lower bound the MSE of the MUSIC algorithm in the \textit{a priori} performance region.
As analyzed, the ZZB converges to the APB in the \textit{a priori} performance region.
Hence, the ZZB provides a global lower bound tighter than the CRB for multiple sources DOA estimation.
Furthermore, the asymptotic region threshold predicted by Fig. \ref{Fig:K5} appears at a lower SNR than that in Fig. \ref{Fig:K1}, which is because the coefficients $2 P_{\mathrm{L}}$ and $\Gamma_{\frac{3}{2}} \! \left( \tilde{u} \right)$ respectively reach $0$ and $1$ more rapidly with $K$ increasing.

\begin{figure*}[t!]
\begin{minipage}[t]{0.49\linewidth}
\centering
\includegraphics[width=1\textwidth]{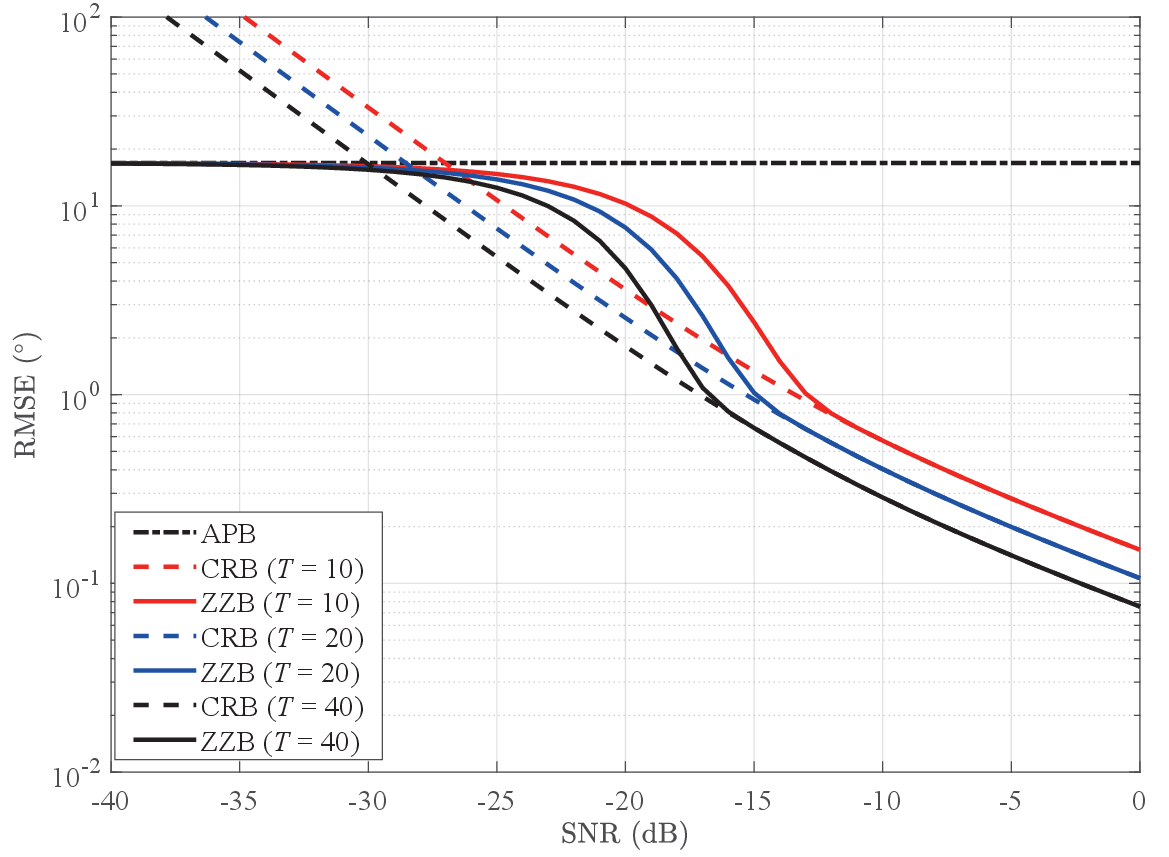}
\caption{Effect of the number of snapshots on ZZB.}\label{Fig:ULA_snap}
\end{minipage}
\hfill
\begin{minipage}[t]{0.49\linewidth}
\centering
\includegraphics[width=1\textwidth]{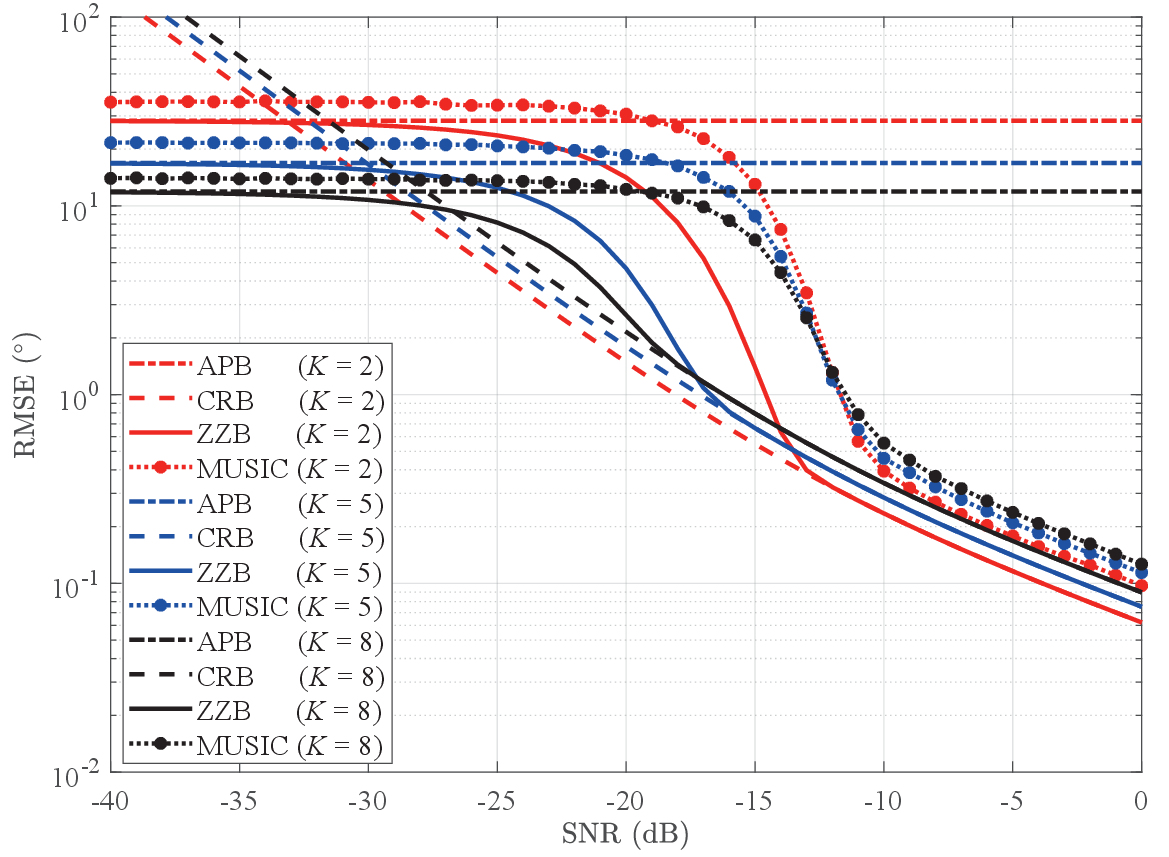}
\caption{Effect of the number of sources on ZZB.}\label{Fig:ULA_sources}
\end{minipage}
\end{figure*}

In Fig. \ref{Fig:ULA_snap},
we investigate the effect of the number of snapshots $T$ on the ZZB, where the number of sources is fixed to $K=5$. It is observed that, the ZZB converges to the APB in the \textit{a priori} performance region regardless of the number of snapshots.
This is because the MSE convergency in the \textit{a priori} performance region only depends on the \textit{a priori} distribution of DOAs and the number of sources. Nevertheless, we prefer more snapshots to bring the lower bound outside the \textit{a priori} performance region, which implies the better estimation performance. Specifically, with the increase of $T$, the ZZB touches the CRB at a lower SNR, leading to a narrower transition region. Each double snapshots brings a fixed decrease of the threshold point, which helps to predict the asymptotic region threshold of the ZZB under different numbers of snapshots.

We then evaluate the effect of the number of sources $K$ on the ZZB in Fig. \ref{Fig:ULA_sources}.
It is observed that, for different numbers of sources $K$, the ZZB always effectively lower bounds the corresponding MSE of the MUSIC algorithm, and respectively converges to the corresponding APB \eqref{ZZB converge low} in the \textit{a priori} performance region and the CRB \eqref{ZZB converge high} in the asymptotic region. With the number of sources $K$ increasing, the ZZB touches the corresponding CRB at a lower SNR, since the function $\Gamma_{\frac{3}{2}}\left( \tilde{u} \right)$ w.r.t. $K$ increases to $1$ (i.e., ZZB converges to CRB) more rapidly. Meanwhile, the ZZB converges to a lower APB in the \textit{a priori} performance region and a higher CRB in the asymptotic region, such that the difference between the ZZB and the corresponding CRB in the transition region becomes smaller. Even so, the ZZB always converges to the APB with SNR decreasing, while the CRB becomes invalid to evaluate the estimation performance of DOA estimator (e.g., MUSIC) in low SNR situation.

\begin{figure*}[t!]
\begin{minipage}[t]{0.49\linewidth}
\centering
\includegraphics[width=0.99\textwidth]{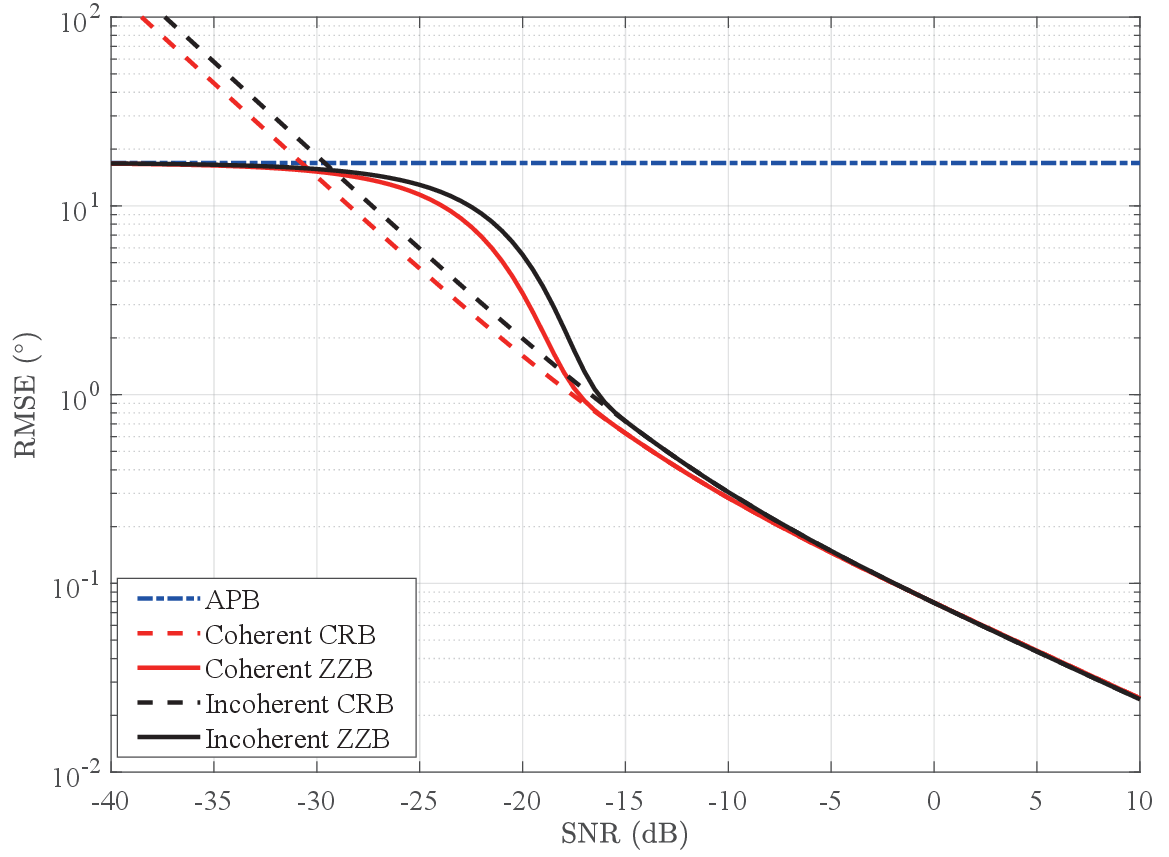}
\caption{Effect of the mutual coherence on ZZB.}\label{bound_coherent}
\end{minipage}
\hspace{2pt}
\hfill
\begin{minipage}[t]{0.49\linewidth}
\centering
\includegraphics[width=0.99\textwidth]{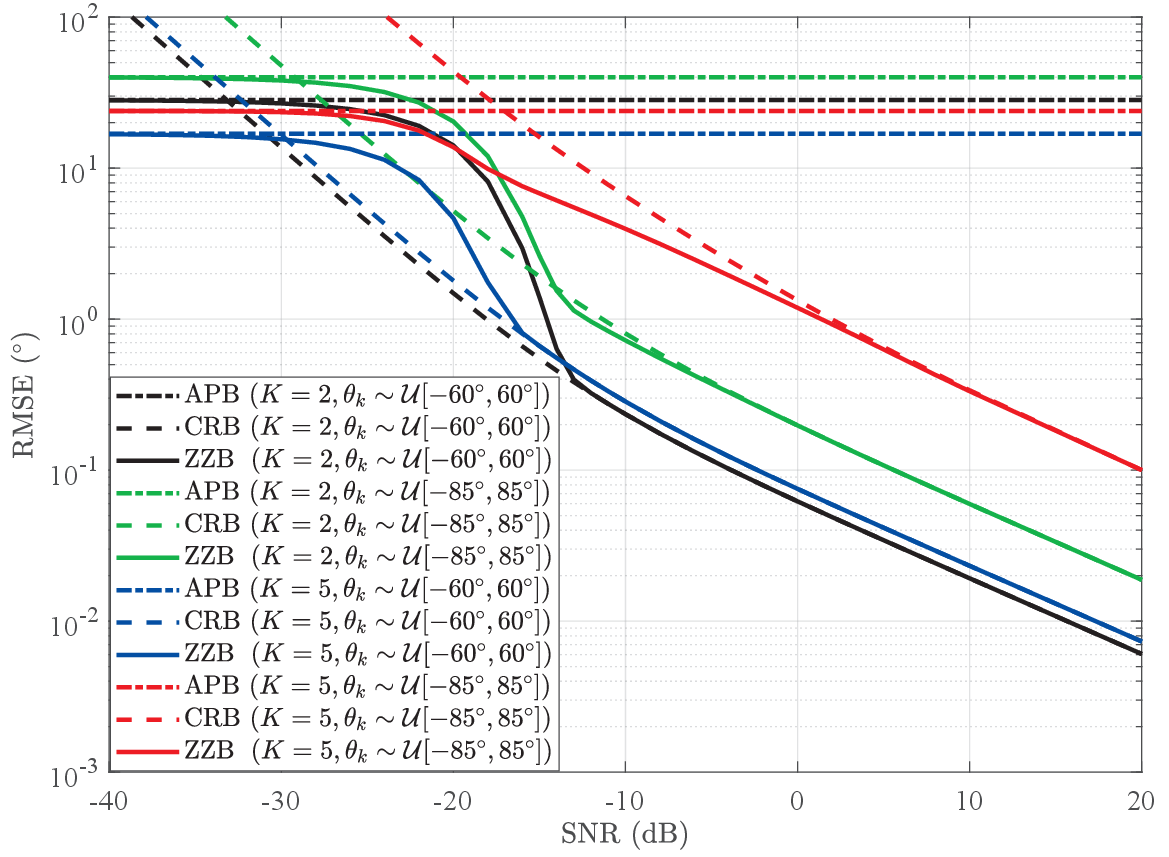}
\caption{Effect of the \textit{a priori} distribution of DOAs on ZZB.}\label{Fig:ULA_range}
\end{minipage}
\end{figure*}

Now, we evaluate the ZZB for the hybrid coherent/incoherent multiple sources DOA estimation. We assume there are $K=5$ sources, where the first $L=3$ sources are mutually coherent with the coefficient $\bm{\beta} = [1,0.9e^{j\phi_1},0.8e^{j\phi_2}]$, whereas the last $K-L=2$ sources are incoherent. Here, both $\phi_1$ and $\phi_2$ are random phases following a uniform distribution $\mathcal{U}[-\pi,\pi]$.
In addition, we also plot the fully incoherent ZZB as a reference.
For fair comparison, the SNRs of $5$ sources in the fully incoherent case are
$\eta_2 = |\beta_2|^2 \eta_1$, $\eta_3 = |\beta_3|^2 \eta_1$, $\eta_4=\eta_5=\eta_1$, respectively.
It is observed from Fig. \ref{bound_coherent} that, both coherent ZZB and incoherent ZZB converge to the APB in the \textit{a priori} performance region, which indicates that, whether coherent or not does not affect the ZZB there.
On the other hand, coherent ZZB and incoherent ZZB converge to their corresponding CRBs in the asymptotic region. However, the coherent ZZB is lower than the incoherent ZZB in the transition region. The reason is that, the ZZB is a linear combination between the APB and the CRB, where the coefficient
$2P_{\mathrm{L}}$ in the coherent ZZB decreases from $1$ more rapidly than that in the incoherent ZZB (as shown in Fig. \ref{Fig:w coherent}), making the coherent ZZB decreases from the APB faster. Thus, the coherent ZZB is lower than the incoherent ZZB.

Then, we investigate the effect of the \textit{a priori} distribution of DOAs on the ZZB for different numbers of sources. Considering that linear array does not perform well for DOAs estimation at directions close to the array endfires, actual array processing systems usually select a narrow field of view. It can be explained from the perspective of performance analysis that, the Fisher information matrix $\bm{J}$ at endfire DOAs close to $\pm 90^\circ$ becomes ill-conditioned, which makes the CRB invalid even in high SNR scenario.
It is observed from Fig. \ref{Fig:ULA_range} that, for the same number of sources $K$, the wider \textit{a priori} distribution of DOAs $\mathcal{U}[-85^\circ, 85^\circ]$ leads to the larger ZZB due to the larger CRB and APB. On the other hand, with the number of sources increasing, the gap of ZZBs between the narrow \textit{a priori} distribution of DOAs $\mathcal{U}[-60^\circ, 60^\circ]$ and the wide \textit{a priori} distribution of DOAs $\mathcal{U}[-85^\circ, 85^\circ]$ becomes larger in the asymptotic region. The reason is that, it is of higher probability that more signals impinge from outside of $[-60^\circ, 60^\circ]$ simultaneously, which makes the estimation accuracy even worse due to the smaller effective array aperture.

\begin{figure}[!t]
\centering
\includegraphics[width=0.49\textwidth]{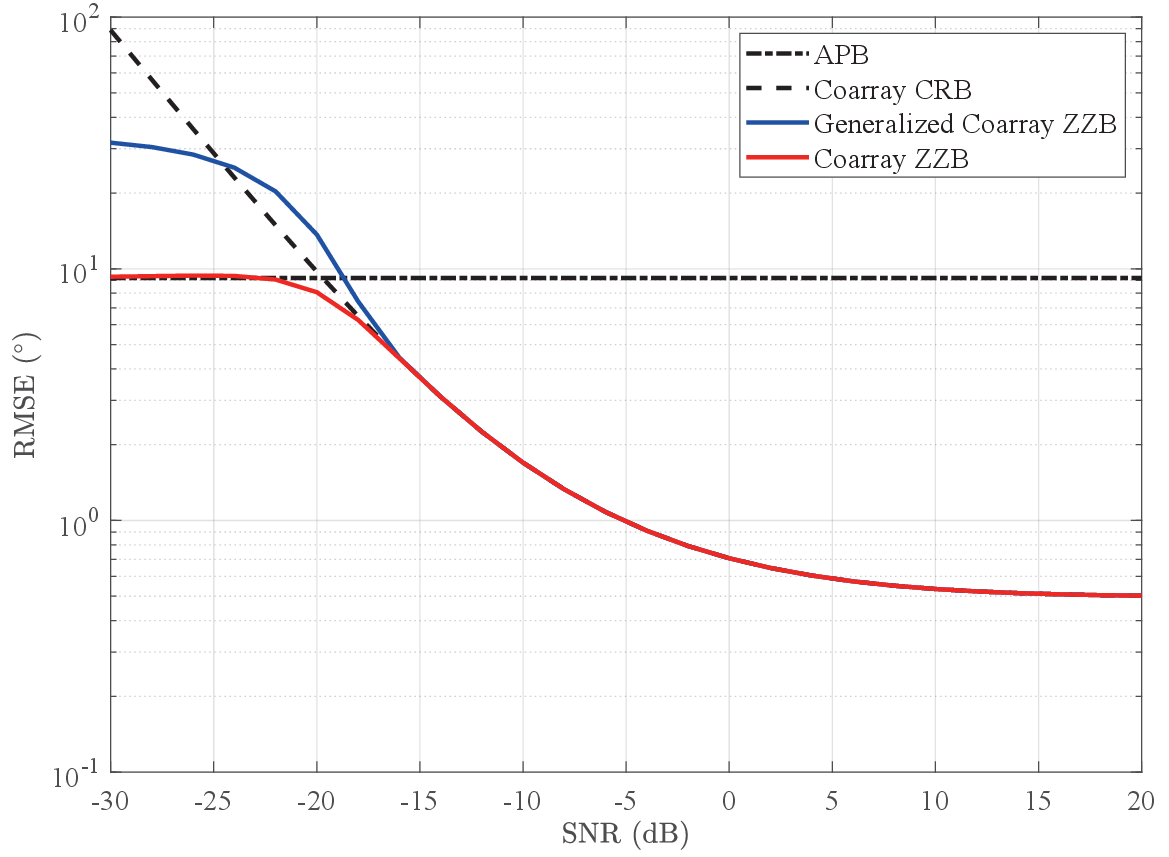}
\vspace{-6pt}
\caption{ZZB for underdetermined DOA estimation.}\label{Fig:CLA}
\vspace{-6pt}
\end{figure}

In addition to the overdetermined DOA estimation, the ZZB is also appropriate for the underdetermined DOA estimation by incorporating the coarray Fisher information \eqref{Coarray FIM}. Here, we adopt a $(3,5)$ pair coprime linear array (the number of array sensors $M=2\times3+5-1=10$) \cite{pal2011coprime, ZHANG2019AnIDFT} to estimate $K=11$ sources following the same \textit{a priori} distribution $\mathcal{U}[-60^{\circ},60^{\circ}]$ with the minimum interval of $5^{\circ}$. Both the coarray CRB \cite[Eq. 37]{Liu2017Cramer} and the generalized coarray ZZB are plotted for a comparison. All the sources are assumed to be fully incoherent with the same SNR. It is observed from Fig. \ref{Fig:CLA} that, the coarray ZZB converges to the coarray CRB in the asymptotic region, and also flattens out as analyzed when the SNR gets higher.
On the other hand, the coarray ZZB converges to the APB as predicted in the \textit{a priori} performance region. Hence, the ZZB also provides a unified bound for underdetermined DOA estimation from the \textit{a priori} performance region to the asymptotic region.

\section{Conclusions}\label{Conclusions}
In this paper, we derived an explicit ZZB for hybrid coherent/incoherent multiple sources DOA estimation.
Different from the single source DOA estimation, the ordering process for eliminating permutation ambiguity during MSE calculation of multiple sources DOA estimation makes the generalized ZZB invalid outside the asymptotic region. Hence, we for the first time introduced the order statistics to describe the effect of the ordering process on the ZZB, which kept it global tight for evaluating multiple sources DOA estimation. The derived ZZB for the first time revealed the relationship between the MSE convergency and the number of sources, regardless of the number of snapshots and array configuration.
Simulation results demonstrate the advantages of the ZZB over the CRB in a wide range of SNR for multiple sources DOA estimation. It is also noted that, in the transition region, the ZZB may be affected by Fisher information when two sources are closely located, which will be further studied in our future work. Finally, we believe the basic conclusions observed in this paper are valid for other multiple sources parameter estimation problems with permutation ambiguity including range, velocity, and so on.


\appendices
\section{Derivation of $\mu(p;\bm{\delta})|_{p=\frac{1}{2}}$ and $\frac {\partial^2 \mu(p;\bm{\delta})} {\partial p^2}\big|_{p=\frac{1}{2}}$}\label{AppA}
In this appendix, we generalize the derivation of $\mu(p;\bm{\delta})|_{p=\frac{1}{2}}$ and $\frac {\partial^2 \mu(p;\bm{\delta})} {\partial p^2}\big|_{p=\frac{1}{2}}$ for single source in \cite{Bell1996Explicit} to hybrid coherent/incoherent multiple sources case.
For the observation data matrix $\bm{X}$ with $T$ snapshots, $\mu(p;\bm{\delta})$ is explicitly expressed as \cite[p. 67]{van2001detection}
\begin{equation}\label{mu reduce}
\begin{aligned}
 \mu(p;\bm{\delta}) \!=\!& \ T \big[ p \ln | \bm{R}_{\bm{x}|\bm{\varphi}} | + (1 - p) \ln | \bm{R}_{\bm{x}|\bm{\varphi}+\bm{\delta}} | \\
 &\quad - \ln | p \bm{R}_{\bm{x}|\bm{\varphi}} + (1 - p) \bm{R}_{\bm{x}|\bm{\varphi}+\bm{\delta}}  | \big],
\end{aligned}
\end{equation}
where $\bm{R}_{\bm{x}|\bm{\varphi}} \triangleq \bm{R}_{\bm{x}|\bm{\theta}=\bm{\varphi}}$ and $\bm{R}_{\bm{x}|\bm{\varphi} + \bm{\delta}} \triangleq \bm{R}_{\bm{x}|\bm{\theta}=\bm{\varphi} + \bm{\delta}}$.

The first derivative of $\mu(p;\bm{\delta})$ w.r.t. $p$ is
\begin{equation}\label{mu first derivative}
\begin{aligned}
     \!\!  \frac {\partial \mu(p; \!\bm{\delta})} {\partial p} & \!= \! T  \Big[\! \ln | \bm{R}_{\bm{x}|\bm{\varphi}} |  \!- \!  \ln | \bm{R}_{\bm{x}|\bm{\varphi}+\bm{\delta}} | \\
 &  \quad\quad - \! \mathrm{Tr} \! \left\{ \!\left[ p \bm{R}_{\bm{x}|\bm{\varphi}}  \!\!+ \! (1 \!- \!p) \bm{R}_{\bm{x}|\bm{\varphi}+\bm{\delta}} \right]^{\!- \!1} \! \! \bm{R}_{-} \! \right\} \!\! \Big]
\end{aligned}
\end{equation}
by denoting $\bm{R}_{-} = \bm{R}_{\bm{x}|\bm{\varphi}} - \bm{R}_{\bm{x}|\bm{\varphi}+\bm{\delta}}$, and then the second derivative of $\mu(p;\bm{\delta})$ w.r.t. $p$ is
\begin{equation}\label{mu second derivative}
\begin{aligned}
 \!\!\!\! \frac {\partial^2\!\mu(p;\!\bm{\delta})} {\partial p^2}
 \!=\! T \mathrm{Tr} \left\{\!  \left[  \left[ p \bm{R}_{\bm{x}|\bm{\varphi}} + (1\!-\!p) \bm{R}_{\bm{x}|\bm{\varphi}+\bm{\delta}} \right]^{\!-\!1} \!\! \bm{R}_{-} \right]^{\!2} \right\}.
\end{aligned}
\end{equation}
Hence, by denoting $\bm{R}_{+} = \bm{R}_{\bm{x}|\bm{\varphi}} + \bm{R}_{\bm{x}|\bm{\varphi}+\bm{\delta}} $, $\mu(p;\bm{\delta})|_{p=\frac{1}{2}}$ and $\frac {\partial^2 \mu(p;\bm{\delta})} {\partial p^2}\big|_{p=\frac{1}{2}}$ respectively become
\begin{equation}\label{mu1/2 reduce}
 \mu(p;\!\bm{\delta})|_{p=\frac{1}{2}} \!=\! T \! \left[\frac{\ln\! \left( | \bm{R}_{\bm{x}|\bm{\varphi}} | | \bm{R}_{\bm{x}|\bm{\varphi}+\bm{\delta}} | \right)   }{2}  \!-\! \ln\! \left| \frac{\bm{R}_{+} }{2}  \right| \right]\!,
\end{equation}
and
\begin{equation}\label{mu1/2 second derivative}
\begin{aligned}
 \frac {\partial^2 \mu(p;\bm{\delta})} {\partial p^2}\bigg|_{p=\frac{1}{2}} =  4 T \mathrm{Tr}\left\{ \left( \bm{R}_{+}^{-1}  \bm{R}_{-}  \right)^{2}  \right\}.
\end{aligned}
\end{equation}

By performing the second-order Taylor series expansion of $\mu(p;\bm{\delta})|_{p=\frac{1}{2}}$ and $\frac {\partial^2 \mu(p;\bm{\delta})} {\partial p^2}\big|_{p=\frac{1}{2}}$ around the point $\bm{\delta}=\bm{0}_{K}$, we have
\begin{equation}\label{Taylor mu1/2 reduce 2}
 \mu(p;\bm{\delta})|_{p=\frac{1}{2}} \approx - \frac{1}{8} \bm{\delta}^\mathrm{T} \bm{J} \bm{\delta},
\end{equation}
and
\begin{equation}\label{Taylor mu1/2 second derivative 2}
\begin{aligned}
 \frac {\partial^2 \mu(p;\bm{\delta})} {\partial p^2}\Bigg|_{p=\frac{1}{2}} \approx  \bm{\delta}^\mathrm{T} \bm{J} \bm{\delta}.
\end{aligned}
\end{equation}
Thus, by substituting \eqref{Taylor mu1/2 reduce 2} and \eqref{Taylor mu1/2 second derivative 2} into \eqref{Pmin}, $P(\bm{\delta})$ is approximated as $P_\mathrm{S}(\bm{\delta})$ given in \eqref{Ps}.
Although the result is consistent with \cite[Eq. (D. 12)]{Bell1996Explicit}, the derivation here is generalized to multiple sources DOA estimation as analyzed below \eqref{Ps}.

Since the approximation $P(\bm{\delta}) \approx P_\mathrm{S}(\bm{\delta})$ is only appropriate for $\bm{\delta}$ located in a small region $\Delta$ around $\bm{0}_{K}$, we further investigate \eqref{Pmin} for $\bm{\delta}$ outside the region $\Delta$. According to Sylvester's determinant theorem
\begin{equation}\label{determinant identity}
 | \bm{I}_{M} + \bm{U}\bm{V} | = | \bm{I}_{K} + \bm{V}\bm{U} |
\end{equation}
for $\bm{U} \in \mathbb{C}^{M \times K}$ and $\bm{V} \in \mathbb{C}^{K \times M}$, the determinants in \eqref{mu1/2 reduce} can be respectively written as
\begin{equation}\label{det example}
\begin{aligned}
    | \bm{R}_{\bm{x}|\bm{\varphi}} |
    = \sigma_n^{2M} \left| \bm{I}_{K} + \frac{1}{\sigma_n^2}  \bm{A}^\mathrm{H} (\bm{\varphi}) \bm{A}(\bm{\varphi}) \bm{\Sigma} \right| ,
\end{aligned}
\end{equation}
\begin{equation}\label{det example2}
\begin{aligned}
   \!\!\!\! | \bm{R}_{\bm{x}|\bm{\varphi}+\bm{\delta}} |
    \!=\! \sigma_n^{2M} \! \left| \bm{I}_{K} \!+\! \frac{1}{\sigma_n^2}  \bm{A}^\mathrm{\!H} (\bm{\varphi} \!+\! \bm{\delta}) \bm{A}(\bm{\varphi}\!+\!\bm{\delta}) \bm{\Sigma} \right| \! ,
\end{aligned}
\end{equation}
and
\begin{equation}\label{det example3}
\begin{aligned}
    \bigg| \frac{1}{2} \bm{R}_{+} \bigg|
    = \sigma_n^{2M} \left| \bm{I}_{2K} + \frac{1}{2\sigma_n^2} \bm{B}^\mathrm{H} \bm{B} \bm{D}  \right|,
\end{aligned}
\end{equation}
where
\begin{equation}\label{X}
\begin{aligned}
    \bm{B} = [ \bm{A}(\bm{\varphi}), \bm{A}(\bm{\varphi} + \bm{\delta})],
\end{aligned}
\end{equation}
and
\begin{equation}\label{D}
\begin{aligned}
    \bm{D} = \mathrm{block \ diag}[\bm{\Sigma}_{co},\bm{\Sigma}_{in},\bm{\Sigma}_{co},\bm{\Sigma}_{in}].
\end{aligned}
\end{equation}
Obviously, $\bm{A}^\mathrm{H} (\bm{\varphi}) \bm{A}(\bm{\varphi})$, $\bm{A}^\mathrm{H} (\bm{\varphi}+\bm{\delta}) \bm{A}(\bm{\varphi}+\bm{\delta})$, and $\bm{B}^\mathrm{H} \bm{B}$ are related to the array beampattern, whose null points coincide with the minima of $P(\bm{\delta})$. The null points of the array beampattern occur at
\begin{equation}\label{single source large delta}
 \left|\bm{a}^\mathrm{H}(\theta) \bm{a}(\theta+\delta) \right| = 0
\end{equation}
for single source case \cite{Bell1996Explicit}, which implies the steering vector $\bm{a}(\theta+\delta)$ is orthogonal to $\bm{a}(\theta)$. By generalizing \eqref{single source large delta} for multiple sources case, we propose the following approximation
\begin{equation}\label{Equivalent approx A}
 \bm{A}^\mathrm{H}(\bm{\theta}) \bm{A}(\bm{\theta}) \approx M \bm{I}_{K}
\vspace{-2pt}
\end{equation}
by considering that multiple steering vectors are approximately orthogonal to each other, from which we have\footnote{To obtain the minimum value of $P(\bm{\delta})$, we only consider the case there is no same value in $\bm{\varphi}$ and $\bm{\varphi}+\bm{\delta}$, such that the steering vectors in $\bm{A}(\bm{\varphi})$ are also orthogonal to those $\bm{A}(\bm{\varphi}+\bm{\delta})$.}
\begin{equation}\label{Equivalent approx}
 \bm{B}^\mathrm{H} \bm{B} \approx M \bm{I}_{2K}.
\end{equation}
As such, \eqref{mu1/2 reduce} can be approximated as
\begin{equation}\label{mu1/2 reduce 2}
\begin{aligned}
  \mu(  p  ;\bm{\delta})|_{p=\frac{1}{2}}
  &\approx  T \ln \frac{ \left| \bm{I}_{K} + \frac{M}{\sigma_n^2}  \bm{\Sigma} \right| }{ \left| \bm{I}_{2K} + \frac{M}{2\sigma_n^2} \bm{D}  \right|  }   \\
 &  = T \ln \frac{ 4(1\!+\!M \|\bm{\beta}\|_2^2 \eta_1 ) }{ (2\!+\!M \|\bm{\beta}\|_2^2 \eta_1 )^2 } \!+\! T \!\!\! \sum_{k=L+1}^{K}\!\! \ln \! \frac{4(1\!+\!M\eta_{k})}{(2\!+\!M\eta_{k})^2}.
\end{aligned}
\end{equation}
Then, $\bm{R}_{+}$ and $\bm{R}_{-}$ in \eqref{mu1/2 second derivative} can be further written as
\begin{equation}\label{sum two covariance matrix}
\begin{aligned}
  \bm{R}_{+} =  2\sigma_n^2 \bm{I}_{M} + \bm{B} \bm{D} \bm{B}^\mathrm{H},
\end{aligned}
\end{equation}
and
\begin{equation}\label{diff two covariance matrix}
\vspace{-1pt}
\begin{aligned}
  \bm{R}_{-} = \bm{B} \tilde{\bm{D}} \bm{B}^\mathrm{H},
\end{aligned}
\vspace{-1pt}
\end{equation}
respectively, where
\begin{equation}\label{P-}
\vspace{-1pt}
\begin{aligned}
    \tilde{\bm{D}} = \mathrm{block \ diag}[\bm{\Sigma}_{co},\bm{\Sigma}_{in},-\bm{\Sigma}_{co},-\bm{\Sigma}_{in}].
\end{aligned}
\vspace{-1pt}
\end{equation}

According to Woodbury matrix identity, the inverse of $\bm{R}_{+}$ can be approximated as
\begin{equation}\label{inverse of sum}
\vspace{-1pt}
\begin{aligned}
   \bm{R}_{+} ^{-1}  \approx \frac{1}{2\sigma_n^2} \bm{I}_{M} - \frac{1}{4\sigma_n^4} \bm{B} \bm{D}  \bm{Y} \bm{B}^\mathrm{H}
\end{aligned}
\vspace{-1pt}
\end{equation}
with the approximation condition \eqref{Equivalent approx}, where
\begin{equation}\label{Y origin}
\begin{aligned}
   \bm{Y} & = \left( \bm{I}_{2K} + \frac{M}{2\sigma_n^2} \bm{D}  \right)^{-1} \\
          & = \mathrm{block \ diag}[\bm{Y}_{co},\bm{Y}_{in},\bm{Y}_{co},\bm{Y}_{in}]
\end{aligned}
\end{equation}
with
\begin{equation}\label{Yc}
\begin{aligned}
    \bm{Y}_{co} & = \left(\bm{I}_{L} +  \frac{M}{2\sigma_n^2} \bm{\Sigma}_{co}\right)^{-1}\\
                & = \bm{I}_{L} - \frac{M\eta_{1}}{ 2 + M \|\bm{\beta}\|_{2}^{2} \eta_{1} }  \bm{\beta}\bm{\beta}^{\mathrm{H}}
\end{aligned}
\end{equation}
according to Sherman-Morrison formula, and
\begin{equation}\label{Yu}
\begin{aligned}
    \!\!\!\!\!\!\bm{Y}_{in} & =  \left(\bm{I}_{K - L}  +  \frac{M}{2\sigma_n^2}\bm{\Sigma}_{in} \right)^{-1} \\
    & = \mathrm{diag}\left[\left(1\!+\!\frac{M\eta_{L+1}}{2}\right)^{\!-1}\!\!,\cdots,\left(1\!+\!\frac{M\eta_{K}}{2}\right)^{\!-1}\right]\!.
\end{aligned}
\end{equation}
Accordingly, we have
\begin{equation}\label{sqaure}
\begin{aligned}
  \!\!\!\!\! & ( \bm{R}_{+}^{-1} \! \bm{R}_{-} )^2 \\
     \!\!\!\!\! & \approx
  \frac{M}{4\sigma_n^4}\Big( \bm{B} \bm{D}^2 \bm{B}^\mathrm{H}
  \!\!-\! \frac{M}{\sigma_n^2} \bm{B}  \bm{D}^{3} \bm{Y} \bm{B}^\mathrm{H}\!\!
  + \frac{M^2}{4\sigma_n^4} \bm{B} \bm{D}^4 \bm{Y}^2 \bm{B}^\mathrm{H} \Big)
\end{aligned}
\end{equation}
since
\begin{equation}
\bm{D}^2 = \tilde{\bm{D}}^2.
\end{equation}
Then, \eqref{mu1/2 second derivative} is approximated as
\begin{equation}\label{trace}
\begin{aligned}
 &\frac {\partial^2 \mu(p;\bm{\delta})} {\partial p^2}  \bigg|_{p=\frac{1}{2}}
  \!\!\!\!\approx
    \frac{T\!M^2}{\sigma_n^4} \mathrm{Tr} \! \left\{ \! \left( \! \bm{D}^{2}
  \!-\! \frac{M}{\sigma_n^2}  \bm{D}^{3} \bm{Y}
  \!+\! \frac{M^2}{4\sigma_n^4}  \bm{D}^{4} \bm{Y}^{2} \! \right) \! \right\},
\end{aligned}
\end{equation}
where the trace of $\bm{D}^{2}$, $\bm{D}^{3}\bm{Y}$, and $\bm{D}^{4}\bm{Y}^2$ are respectively expressed as
\begin{equation}\label{trace D2}
\begin{aligned}
  \mathrm{Tr}\{ \bm{D}^2 \} = 2 \|\bm{\beta}\|_{2}^{4} \sigma_{1}^{4} +  \!\! \sum_{k=L+1}^{K} 2  \sigma_{k}^4,
\end{aligned}
\end{equation}
\begin{equation}\label{trace D3Y}
\begin{aligned}
  \mathrm{Tr}\{ \bm{D}^{3}\bm{Y} \} =  \frac{4 \|\bm{\beta}\|_2^6 \sigma_{1}^{6} }{2+M \|\bm{\beta}\|_2^2 \eta_1} +  \sum_{k=L+1}^{K} \frac{4\sigma_{k}^{6}}{2+M\eta_k}.
\end{aligned}
\end{equation}
and
\begin{equation}\label{trace D4Y2}
\begin{aligned}
  \mathrm{Tr}\{ \bm{D}^{4}\bm{Y}^{2} \}
                            =  \frac{8 \|\bm{\beta}\|_2^{8} \sigma_{1}^8}{\left(2\!+\!M \|\bm{\beta}\|_2^2 \eta_1\right)^2}\!+ \!\!\!\sum_{k=L + 1}^{K}\!  \frac{8\sigma_{k}^{8}}{(2\!+\!M\eta_l)^2}.
\end{aligned}
\end{equation}
Finally, \eqref{trace} becomes
\begin{equation}\label{trace 2}
\begin{aligned}
 \frac {\partial^2 \mu(p;\bm{\delta})} {\partial p^2}  \bigg|_{p=\frac{1}{2}}
 \!\!\!\! \approx
 8T\left[ \! \left( \frac{M \|\bm{\beta}\|_2^2 \eta_1}{2\!+\!M \|\bm{\beta}\|_2^2 \eta_1} \right)^{\!2}\!\!\! + \!\!\!\sum_{k=L\!+\!1}^{K} \!\! \left(\frac{M\eta_k}{2\!+\!M\eta_k}\right)^{\!2}\right]\!,
\end{aligned}
\end{equation}
which is a function of SNR $\eta_k$, the number of array sensors $M$, the number of sources $K$ and the coherent coefficient $\bm{\beta}$.
By substituting \eqref{mu1/2 reduce 2} and \eqref{trace 2} into \eqref{Pmin}, $P(\bm{\delta})$ for $\bm{\delta}$ outside the region $\Delta$ is approximated as $P_{\mathrm{L}}$, which is given in \eqref{PL}.

The boundary of the region $\Delta$ occurs at $P_{\mathrm{S}}(\bm{\delta}) = P_{\mathrm{L}}$. However, the explicit boundary is difficult to calculate. Similar to \cite{Bell1996Explicit}, considering that $\mathcal{Q}(z)$ is a decreasing function w.r.t. $z$, the region $\Delta$ can be approximately given by
\begin{equation}\label{alt Threshold}
\frac{\sqrt{\bm{\delta}^\mathrm{T} \bm{J} \bm{\delta}}}{2} \leq \! \sqrt{2T \left[ \left( \frac{M \|\bm{\beta}\|_2^2 \eta_1}{2\!+\!M \|\bm{\beta}\|_2^2 \eta_1} \right)^2 \!+\! \!\sum_{k=L\!+\!1}^{K} \!  \left(\frac{M\eta_k}{2\!+\!M\eta_k}\right)^{\!2}\right]},
\end{equation}
which is the interior of an ellipse given in \eqref{small region}.

\section{Derivation of \eqref{ZZB final}}\label{AppB}
In this appendix, we follow the framework in \cite{Bell1996Explicit} to solve the optimization problem in \eqref{Eq:ZZB vector same weight}. It is worth noting that, in our derivation, $\bm{\theta}$ follows uniform \textit{a priori} distribution defined in the $K$-dimensional space rather than defined on a subset of the unit disc in \cite{Bell1996Explicit}. First, by substituting \eqref{Pmin final} into \eqref{K dimensional integration}, we have
\begin{equation}\label{Eq:AP}
\begin{aligned}
 &      \frac{P(\bm{\delta})}{\zeta^K} \prod\limits_{k=1}^{K}   (\zeta - |\delta_k|) \\
 &    \approx     \frac{P_{\mathrm{L}}}{\zeta^K} \prod\limits_{k=1}^{K}  (\zeta - |\delta_k|)
    +  \left\{
    \begin{array}{lc}
        P_{\mathrm{S}}(\bm{\delta})  -  P_{\mathrm{L}} &    \bm{\delta} \in \Delta \\
        0 &    \bm{\delta} \notin \Delta\\
    \end{array}
,\right.
\end{aligned}
\vspace{-2pt}
\end{equation}
based on which the corresponding optimization problem in \eqref{Eq:ZZB vector same weight} can be divided into three separated terms as
\begin{equation}\label{Eq:max AP}
\begin{aligned}
& \max \limits_{ \bm{\delta} : \bm{1}_{K}^{\mathrm{T}}\bm{\delta} = \sqrt{K} h }   \frac{P(\bm{\delta})}{\zeta^K} \prod\limits_{k=1}^{K} (\zeta-|\delta_k|)\\
& \approx \!\!\!\!\!\!\!  \max \limits_{\bm{\delta} : \bm{1}_{K}^{\mathrm{T}}\bm{\delta} = \sqrt{K} h }  \frac{ P_{\mathrm{L}} }{\zeta^K} \prod\limits_{k=1}^{K} (\zeta-|\delta_k|)
+ \!\!\!\!\!\!\! \max \limits_{ \bm{\delta} \in \Delta :   \bm{1}_{K}^{\mathrm{T}}\bm{\delta} = \sqrt{K} h } \!\!\!\!\!\!\! P_{\mathrm{S}}(\bm{\delta}) - P_{\mathrm{L}}.
\end{aligned}
\end{equation}
Then, we have the following property about the maximum of $\frac{1}{\zeta^K} \prod\limits_{k=1}^{K} (\zeta-|\delta_k|)$.
\newtheorem{property}{Property}
\begin{property} \label{maximum solution}
\begin{equation}
\vspace{-2pt}
\begin{aligned}
\max \limits_{ \bm{\delta} : \bm{1}_{K}^{\mathrm{T}}\bm{\delta} = \sqrt{K} h } \frac{1}{\zeta^K} \prod\limits_{k=1}^{K} (\zeta-|\delta_k|)
 =  \left( 1 - \frac{h}{\zeta\sqrt{K}}\right)^K .
\end{aligned}
\end{equation}
\end{property}
\begin{IEEEproof} \label{maximum solution proof}
Obviously, the integration region $\Phi$ in \eqref{Eq:Phi region} brings the constraint
\begin{equation}\label{constraint2}
|\delta_k| \leq \zeta , \forall k=1,2,\cdots,K,
\end{equation}
such that all the $\zeta-|\delta_k|$ must be positive.
With the constraint
\begin{equation}\label{constraint}
\bm{1}_{K}^{\mathrm{T}}\bm{\delta} =\sqrt{K}h,
\end{equation}
the maximum of $\prod_{k=1}^{K} (\zeta-|\delta_k|)$ must occur in the region $\Omega = \{\delta_k>0, \forall k=1,2,\cdots,K\}$, because
given any $\bm{\delta}$ containing negative elements, there always exists a certain $\bm{\delta}$ in the region $\delta^{+}$ leading to a larger $\prod_{k=1}^{K} (\zeta-|\delta_k|)$. Thus, we have
\begin{equation}
 \sum_{k=1}^{K} \zeta-|\delta_k| = \sum_{k=1}^{K} \zeta-\delta_k = K\zeta -\sqrt{K}h,
\end{equation}
which is a positive constant. Based on the arithmetic-geometric mean inequality, the maximum of $\prod_{k=1}^{K} (\zeta\!-\!|\delta_k|)$ occurs if and only if $\delta_k = \frac{h}{\sqrt{K}}, \ \forall k = 1,2,\cdots,K$, and thus Property \ref{maximum solution} is proved.
\end{IEEEproof}

Accordingly, the first term in \eqref{Eq:max AP} becomes
\begin{equation}\label{Eq:max 1}
\begin{aligned}
\max \limits_{ \bm{\delta} : \bm{1}_{K}^{\mathrm{T}}\bm{\delta} = \sqrt{K} h }  \frac{P_{\mathrm{L}} }{\zeta^K} \prod\limits_{k=1}^{K}  (\zeta - |\delta_k|)
 =  P_{\mathrm{L}} \left( 1 - \frac{h}{\sqrt{K}\zeta} \right)^{ K}.
\end{aligned}
\end{equation}
The maximum of the second term  in \eqref{Eq:max AP} occurs at \cite{Bell1996Explicit}\footnote{Since the array observation data are related to the wanted random parameters (i.e., DOAs) and other unwanted random parameters (e.g., sources and noise power), we cannot simply calculate $\bm{J}^{-1}$ only considering the Fisher information w.r.t. $\bm{\theta}$ \cite[Chap. 8]{van2002detection}. In this case, the parameter vector containing all random parameters is $\bm{\alpha} = [ \bm{\theta}^{\mathrm{T}}, \bm{\sigma}^{\mathrm{T}}]^{\mathrm{T}}$ with $\bm{\sigma} = [\sigma_{1}^2,\sigma_{2}^2,\cdots,\sigma_{K}^2,\sigma_n^2]^{\mathrm{T}}$ denoting the unwanted parameter vector, and the complete Fisher information matrix $\bm{J}_{\bm{\alpha}}$ can be calculated according to \eqref{Eq:Fisher under} or \eqref{Coarray FIM} by replacing $\bm{\theta}$ into $\bm{\alpha}$. Then, $\bm{J}^{-1}$ should be calculated as the first block with $K \times K$ elements in $\bm{J}_{\bm{\alpha}}^{-1}$, whose explicit solution is given in \cite[Eq. (3.1)]{Stoica1990Performance} for overdetermined DOA estimation and \cite[Eq. (43)]{Liu2017Cramer} for underdetermined DOA estimation, respectively.}
\begin{equation}
\bm{\delta}
=\sqrt{K}h \frac{\bm{J}^{-1}\bm{1}_{K}}{\bm{1}_{K}^{\mathrm{T}} \bm{J}^{-1} \bm{1}_{K} },
\end{equation}
such that it becomes
\begin{equation}\label{Eq:max 2}
\begin{aligned}
 \max \limits_{ \bm{\delta} \in \Delta :  \bm{1}_{K}^{\mathrm{T}}\bm{\delta} = \sqrt{K} h }  P_{\mathrm{S}}(\bm{\delta}) = \mathcal{Q}\left( \frac{\sqrt{K}h}{2\sqrt{\bm{1}_{K}^{\mathrm{T}} \bm{J}^{-1} \bm{1}_{K}} } \right).
\end{aligned}
\end{equation}
Correspondingly, \eqref{Eq:max AP} is further written as
\begin{eqnarray}\label{Eq:max AP 2}
   \max \limits_{ \bm{\delta} : \bm{1}_{K}^{\mathrm{T}}\bm{\delta} = \sqrt{K} h }  \frac{P(\bm{\delta})}{\zeta^K} \prod\limits_{k=1}^{K} (\zeta-|\delta_k|) & \nonumber \\
  &\displaystyle{\!\!\!\!\!\!\!\!\!\!\!\!\!\!\!\!\!\!\!\!\!\!\!\!\!\!\!\!\!\!\!\!\!\!\!\!\!\!\!\! \!\!\!\!\!\!\!\!\!\!\!\!\!\!\!\!\!\!\!\!\!\!\!\!\!\!\!\!\!\!\!\!\!\!\!\!\!\!\!\!\!\! \approx \! P_{\mathrm{L}} \! \left(1\!-\!\frac{h}{\sqrt{K}\zeta}\right)^{\!K}
  \!\!\!\!+\!  \left\{
   \begin{array}{lc}
        \!\!\mathcal{Q}\left( \frac{\sqrt{K}h}{2\sqrt{\bm{1}^{\mathrm{T}} \bm{J}^{-1} \bm{1}} } \right)\!-\!P_{\mathrm{L}}  & 0 \leq h \leq \tilde{h},  \\
        \!\!0 &  h > \tilde{h},  \\
    \end{array}
\right.}
\end{eqnarray}
where $\tilde{h}$ is the threshold derived via \eqref{alt Threshold}, i.e.,
\begin{equation}\label{Eq:hs}
\begin{aligned}
 \frac{\sqrt{K}\tilde{h}}{2\sqrt{\bm{1}_{K}^{\mathrm{T}} \bm{J}^{-1} \bm{1}_{K}} }\approx \!\!\! \sqrt{2T \! \left[ \! \left(\frac{M \|\bm{\beta}\|_2^2 \eta_1}{2\!+\!M \|\bm{\beta}\|_2^2 \eta_1}\right)^{\!2} \!\!\!+\!\!\! \sum_{k=L\!+\!1}^{K} \!\!  \left(\frac{M\eta_k}{2\!+\!M\eta_k}\right)^{\!2}\right]}
\end{aligned}
\end{equation}
with the constraint $ 0 \leq \tilde{h} \leq\sqrt{K}\zeta$.

Then, \eqref{Eq:ZZB vector same weight} can be further written as
\begin{equation}\label{integration wrt h}
\begin{aligned}
  \frac{1}{K} \bm{1}_{K}^\mathrm{T} \bm{R}_{\bm{\epsilon}} \bm{1}_{K}
  & \! \geq \! P_{\mathrm{L}}   \int_{0}^{\sqrt{K}\zeta} \!\! \left(1 \!- \! \frac{h}{\sqrt{K}\zeta}\right)^{\!K} \!\! h dh \\
  & \ \ +    \int_{0}^{\tilde{h}} \! \left[ \! \mathcal{Q}\! \left( \! \frac{\sqrt{K}h}{2\sqrt{\bm{1}_{K}^{\mathrm{T}} \bm{J}^{\!-\!1} \bm{1}_{K}} } \! \right)\!-\!P_{\mathrm{L}}\! \right] \!\! h dh,
\end{aligned}
\end{equation}
where the first integration
\begin{equation}\label{Eq:first integration}
\begin{aligned}
\int_{0}^{\sqrt{K}\zeta}   \left(1\!-\!\frac{h}{\sqrt{K}\zeta}\right)^{\!K}   h dh
=  \frac{12 \bm{1}_{K}^{\mathrm{T}} \bm{R}_{\bm{\theta}} \bm{1}_{K}  }{(K\!+\!1)(K\!+\!2)},
\end{aligned}
\end{equation}
and the second integration
\begin{equation}\label{Eq:second integration}
\begin{aligned}
& \int_{0}^{\tilde{h}} \left[ \mathcal{Q} \left(  \frac{\sqrt{K}h}{2\!\sqrt{\bm{1}_{K}^{\mathrm{T}} \bm{J}^{-\!1} \bm{1}_{K}} } \right)-P_{\mathrm{L}} \right] h dh \\
&= \int_{0}^{\tilde{h}} \frac{1}{\sqrt{2\pi}}e^{-\frac{K h^2}{8\bm{1}_{K}^{\mathrm{T}} \bm{J}^{-1} \bm{1}_{K}}} \frac{\sqrt{K}h^2}{4\sqrt{\bm{1}_{K}^{\mathrm{T}} \bm{J}^{-1} \bm{1}_{K}}}d h \\
& = \Gamma_{\frac{3}{2}}(\tilde{u})  \frac{\bm{1}_{K}^{\mathrm{T}} \bm{J}^{-\!1}  \bm{1}_{K} }{K}.
\end{aligned}
\end{equation}
with $\tilde{u}$ defined in \eqref{us}.



\ifCLASSOPTIONcaptionsoff
  \newpage
\fi



%
%
%
\footnotesize
\bibliography{ZZB_IEEE}

\bibliographystyle{IEEEtran}

\end{document}